\documentclass[twocolumn,superscriptaddress,10pt]{revtex4-2}
\usepackage[T1]{fontenc}
\usepackage{times}
\usepackage[utf8]{inputenc}
\usepackage[resetlabels,labeled]{multibib}
\usepackage{amsmath}
\usepackage{amsthm}
\usepackage{amsfonts}
\usepackage{amssymb}
\usepackage{lipsum} 
\usepackage{bbm}
\usepackage{bbold}
\usepackage{graphicx,epsfig,color,tcolorbox}
\usepackage{physics}
\usepackage{mathtools}
\usepackage{mathrsfs}
\usepackage[colorlinks=true,linkcolor=teal,citecolor=purple]{hyperref}
\usepackage{hyperref}
\usepackage{cleveref}
\usepackage[mathscr]{eucal}

\usepackage{tikz}
\usetikzlibrary{shapes.geometric}
\usetikzlibrary{positioning}

\newcommand{\plb}{\textcolor{black}}
\newcommand{\Lam}{\mathbb{\Lambda}}

\newcommand{\ot}{\otimes}
\newcommand{\id}{\mathrm{id}}

\newcommand{\fsep}{\mathscr{F}_{\rm SEP}}

\newcommand{\q}{\mathscr{R}_{\rm BN}}
\newcommand{\fbn}{\mathscr{F}_{\rm BN}}
\newcommand{\rbn}{\mathscr{R}_{\rm BN}}
\newcommand{\ft}{\mathscr{F}_{\rm T}}
\newcommand{\rt}{\mathscr{R}_{\rm T}}
\newcommand{\robn}{\mathcal{R}_{\rm{BN}}}
\newcommand{\rot}{\mathcal{R}_{\rm{T}}}
\newcommand{\roe}{\mathcal{R}_{\rm{E}}}
\newtheorem{result}{Result}

\newtheorem{definition}{Definition}

\newtheorem{corollary}{Corollary}
\newtheorem{task}{Task}
\begin{document}

\title{The operational significance of the quantum resource theory of Buscemi nonlocality}
\author{Patryk Lipka-Bartosik}
\address{H.H. Wills Physics Laboratory, University of Bristol, Tyndall Avenue, Bristol, BS8 1TL, United Kingdom}

\author{Andr\'es F. Ducuara}
\address{H.H. Wills Physics Laboratory, University of Bristol, Tyndall Avenue, Bristol, BS8 1TL, United Kingdom}
\address{Quantum Engineering Technology Labs, H. H. Wills Physics Laboratory and \\
Department of Electrical \& Electronic Engineering, University of Bristol, BS8 1FD, UK.}
\address{Quantum Engineering Centre for Doctoral Training, \\H. H. Wills Physics Laboratory and Department of Electrical \& Electronic Engineering, University of Bristol, BS8 1FD, UK}

\author{Tom Purves}
\address{H.H. Wills Physics Laboratory, University of Bristol, Tyndall Avenue, Bristol, BS8 1TL, United Kingdom}

\author{Paul Skrzypczyk} 
\address{H.H. Wills Physics Laboratory, University of Bristol, Tyndall Avenue, Bristol, BS8 1TL, United Kingdom}

\date{\today}

\begin{abstract}
Although entanglement is necessary for observing nonlocality in a Bell experiment, there are entangled states which can never be used to demonstrate nonlocal correlations. In a seminal paper \cite{Buscemi2012} F. Buscemi extended the standard Bell experiment by allowing Alice and Bob to be asked quantum, instead of classical, questions. This gives rise to a broader notion of nonlocality, one which can be observed for every entangled state. In this work we study a resource theory of this type of nonlocality referred to as \emph{Buscemi nonlocality}. We propose a geometric quantifier measuring the ability of a given state and local measurements to produce Buscemi nonlocal correlations and  establish its operational significance. In particular, we show that any distributed measurement which can demonstrate Buscemi nonlocal correlations provides strictly better performance than any distributed measurement which does not use entanglement in the task of distributed state discrimination. We also show that the maximal amount of Buscemi nonlocality that can be generated using a given state is precisely equal to its entanglement content. Finally, we prove a quantitative relationship between: Buscemi nonlocality, the ability to perform nonclassical teleportation, and entanglement. Using this relationship we propose new discrimination tasks for which nonclassical teleportation and entanglement lead to an advantage over their classical counterparts.
\end{abstract}

\maketitle

\section{Introduction}

Quantum entanglement is one of the most characteristic features of quantum theory \cite{Horodecki_2009}. During the early years of its development, however, it was recognized mainly as a bizarre property which distinguished it from classical physics. It was due to the discovery of Bell nonlocality \cite{review_NL} and subsequent development of Bell inequalities which allowed this distinction to be formulated quantitatively and to verify the predictions of quantum theory in an experimentally feasible setting. 

Bell nonlocality is today perceived as a phenomenon in its own right and can be defined and tested irrespectively of the underlying theory. In simple terms Bell nonlocality refers to the situation when correlations shared between spatially separated parties cannot be explained as arising from a shared classical resource. The concept of Bell nonlocality is perhaps best understood in terms of a Bell experiment, which is sometimes also called a ``no-signalling game''. In such a game, a referee distributes two physical systems to two spatially separated players, Alice ($\rm A$) and Bob ($\rm B$). Upon receiving their systems, each player is asked a question from a pre-arranged set of questions, labelled $x$ for Alice and $y$ for Bob. Depending on which of the questions was asked, Alice measures her system locally and obtains an outcome $a$. Similarly, based on his own question, Bob measures his share of the system and obtains $b$. The data produced from the experiment can be described using a conditional probability distribution $p(a,b|x,y)$, that is the probability of producing outcomes $a$ and $b$ given the choice of measurements labelled by $x$ and $y$. 

Importantly, not all entangled states can display Bell non-locality \cite{Werner89, Barrett2002, review_EL}. Quantum states can actually demonstrate other forms of nonlocality which are not accessible in a Bell experiment but which may become apparent in different experimental settings. In a seminal work \cite{Buscemi2012} Buscemi generalized Bell's original experiment by allowing the referee to ask ``quantum questions''. This amounts to replacing the original set of classical (and therefore mutually orthogonal) questions $\left\{\ket{x}\right\}$ with a set of quantum states $\left\{\ket{\omega_x}\right\}$ which need not be orthogonal.  
The correlation data $p(a,b|\omega_x, \omega_y)$ obtained in this modified experiment, dubbed \emph{semi-quantum non-signalling games}, differs significantly from its archetypical counterpart. Perhaps the most striking consequence is that the new experiment is powerful enough to reveal the nonlocality of any entangled quantum state, even the nonlocality which would be hidden under a standard Bell test \cite{Buscemi2012}. This semi-quantum approach, also called measurement-device-independent {(MDI)}, has been a fruitful line of investigation during the last decade \cite{MDI1, MDI2, MDI3, MDI4, MDI5, MDI6, MDI7, MDI8, MDI9, MDI12, MDI13, MDI14, MDI15}.

In this work we propose interpreting the correlation data obtained in a semi-quantum non-signalling game as an indicator of a this type of nonlocality which we refer to as \emph{Buscemi nonlocality}. In order to formalize this notion we utilise the framework of Quantum Resource Theories (QRTs)  \cite{HORODECKI2012, Chitambar2019}. This is a set of tools and techniques developed to systematically quantify different properties of quantum systems. QRTs can be classified in terms of \emph{objects} and \emph{resources} studied in a given theory. Classification of QRTs with respect to the object lead to the resource theories of states  \cite{Chitambar2019}, measurements \cite{Skrzypczyk2019, Designolle2019, DS, Guff2019, Oszmaniec2019,Oszmaniec2017}, channels \cite{Theurer2019, Liu2019, Liu2019op, Wilde2013}, and boxes \cite{DS1, DS2, DS3, DS4}. On the other hand, classifying QRTs with respect to the type of the studied resource leads to the resource theories of pure \cite{Nielsen1999} and mixed-state entanglement \cite{Vidal1999}, coherence \cite{Napoli2016}, purity \cite{Horodecki2003, Streltsov2018}, athermality \cite{Janzing2000, Brandao2013, Brandao2015, Horodecki2013, Ng2018, Horodecki2015}, nonlocality \cite{Cavalcanti2016}, asymmetry \cite{Piani2016}, measurement incompatibility \cite{Buscemi2020}, teleportation \cite{Supic2019, Cavalcanti2017}, magic \cite{magic2017}, nonmarkovianity \cite{bhattacharya2018convex, wakakuwa2017operational, an2019quantifying} or nongaussianity \cite{Takagi2018}, amongst many more. Its worth mentioning that although many QRTs use essentially the same mathematical formalism, their physical implications can be genuinely different. Hence the wide applicability of the framework to otherwise unrelated problems is a truly surprising aspect of Nature.

In this work, we focus on the quantum resource theory of Buscemi nonlocality, which is an instance of the resource theory from \cite{DS3,DS4}. The natural object relevant for this theory is a generalized measurement (POVM) performed by spatially-separated parties that do not communicate (distributed measurement). We investigate a geometric measure that quantifies the amount of Buscemi nonlocality contained within a given distributed measurement termed Robustness of Buscemi Nonlocality (RoBN). We then address Buscemi nonlocality as a property of states, by considering the maximal amount of Buscemi nonlocality that can be obtained using a given state by any local set of measurements on Alice's and Bob's side. 

As our first result we show that Buscemi nonlocality has operational significance, by finding an operational task for which Buscemi nonlocality is a natural resource. This also gives rise to a complete family of monotones for this resource theory. Furthermore, we explore how Buscemi nonlocality relates to other types of nonclassical phenomena: nonclassical teleportation \cite{Cavalcanti2017} and entanglement \cite{Horodecki_2009}. In particular, we show that the maximal amount of RoBN which can be achieved when Bob (Alice) is allowed to use any measurement leads to the so-called Robustness of Teleportation (RoT) of a teleportation channel from Alice (Bob) to Bob (Alice). Using this direct link we show that RoBN of a state is precisely equal to its entanglement content expressed by the Robustness of Entanglement (RoE). These results consequently lead to novel operational interpretations of both quantifiers.

The paper is organized as follows. In Sec. \ref{sec2} we cover the relevant formalism, remind the idea of characterizing  nonlocality in terms of non-signalling games and recall the robustness quantifier of Buscemi nonlocality (RoBN). In Sec. \ref{sec3a} we find its operational interpretation in terms of the advantage in the task of distributed state discrimination (DSD). In Sec. \ref{sec3b} we explore the relationship between Buscemi nonlocality and the concepts of nonclassical teleportation and entanglement. Finally, in Sec. \ref{sec3c} we describe a tangential view on RoBN from the perspective of single-shot information theory. We conclude with Sec. \ref{sec4} where we summarize our findings and highlight several open questions. 

\section{Framework}
\label{sec2}
\begin{figure}
    \centering
    
\begin{tikzpicture}[scale=0.6, every node/.style={scale=1.5}]
\draw[ultra thick, rounded corners=15pt] (1.5, 2) rectangle (-3.5, 4) {};
\draw[ultra thick, rounded corners=15pt] (2.5, 2) rectangle (7.5, 4) {};
\draw[ultra thick, rounded corners=15pt] (-0.5, 0) rectangle (4.5, -2) {};

\draw[very thick] (0.5,0) -- (0.5, 2) {};
\draw[very thick] (3.5,0) -- (3.5, 2) {};
\draw[very thick] (-2.5,2) -- (-2.5, -2.5) {};
\draw[very thick] (6.5,2) -- (6.5, -2.5) {};

\draw[very thick] (6.5,2) -- (6.5, -2.5) {};
\draw[very thick] (6.5,2) -- (6.5, -2.5) {};
\draw[very thick, double] (-1, 4) -- (-1, 6) node at (-0.1, 5) [anchor = east] {$a$};
\draw[very thick, double] (5, 4) -- (5, 6) node at (5.8, 5) [anchor = east] {$b$};

\node at (0.5, -3) {$\rm A'$};
\node at (3.5, -3) {$\rm B'$};
\node at (-2.5, -3) {$\rm A$};
\node at (6.5, -3) {$\rm B$};
\node at (-1, 3) {$\mathbb{M}^{\rm A}$};
\node at (5, 3) {$\mathbb{M}^{\rm B}$};
\node at (2, -1) {$\rho$};

\end{tikzpicture}
    \caption{A schematic diagram of a distributed measurement $\mathbb{M}^{\rm AB}$ composed of local measurements for Alice $\mathbb{M}^{\rm A} = \{M_{a}^{\rm AA'}\}$, for Bob $\mathbb{M}^{\rm B} = \{M_{b}^{\rm B'B}\}$ and a state $\rho^{\rm A'B'}$ shared between them. This is the most general type of measurement which Alice and Bob can perform in a distributed scenario which does not allow for communication. } 
    \label{fig1}
\end{figure}
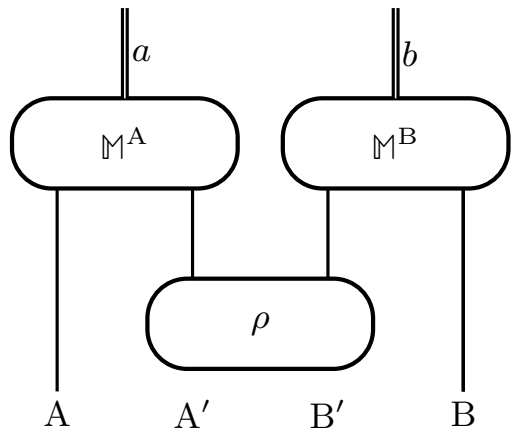

\noindent 
In what follows we will denote a local bipartite measurement on Alice's side (system $\rm AA'$) with $\mathbb{M}^{\rm A} =\{M_{a}^{\rm AA'}\}$, where each $M_{a}^{\rm AA'}$ is a positive semi-definite operator that adds up to the identity (POVM). Similarly we will use $\mathbb{M}^{\rm AB}$ to indicate that the measurement is non-local, i.e. we will treat systems labelled with different letters, e.g. $\rm A$ and $\rm B$ , as two spatially separated parties. We are interested in the most general type of measurement that can be performed in this bipartite scenario without the aid of classical or quantum communication. This can be realized by ($i$) allowing Alice and Bob to apply arbitrary bipartite measurements in their labs, denoted respectively $\mathbb{M}^{\rm A} = \{M_a^{\rm AA'}\}$  and $\mathbb{M}^{\rm B} = \{M_b^{\rm B'B}\}$, where $a \in \{1, \ldots, o_{\rm A}\}$ and $b \in \{1, \ldots, o_{\rm B}\}$ denote Alice's and Bob's outcomes and $(ii)$ allowing the two parties to share a quantum state $\rho^{\rm A'B'}$. In this way Alice and Bob can store and share all types of classical information (e.g. classical memory or measurement strategy), as well as quantum information (i.e. shared entanglement). We denote such a measurement with $\mathbb{M}^{\rm AB} = \{M_{ab}^{\rm AB}\}$, where the corresponding POVM elements are of the following general form:
\begin{align}
    \label{eq:quantum_meas}
    M_{ab}^{\rm AB} = \tr_{\rm A'B'} \! \left[\!\left(M_a^{\rm AA'}\! \ot\! M_b^{\rm B'B}\right)\!\left(\mathbb{1}^{\rm A}\! \ot \rho^{\rm A'B'}\! \ot \mathbb{1}^{\rm B}\right)\!\right].
\end{align}
Since the sets of all quantum states and quantum measurements are both convex sets, it follows that the set of measurements of the form (\ref{eq:quantum_meas}) is also a convex set. We will refer to measurements of the form (\ref{eq:quantum_meas}) as \emph{distributed measurements} and denote the set of all such measurements with $\rbn$. Hence, whenever the elements of $\mathbb{M}^{\rm AB}$ can be written as in (\ref{eq:quantum_meas}) for some choice of shared state and local measurements we will denote it with $\mathbb{M}^{\rm AB} \in \q$. Figure \ref{fig1} illustrates a distributed measurement and describes the relationship between different subsystems. This type of objects appear naturally in a wide range of contexts when studying non-local effects in an MDI setting \cite{Buscemi2012,MDI2,MDI8,MDI10}.

We now specify the most general class of operations that the separated parties in $\rm A$ and $\rm B$ can perform, without communicating, to improve the properties of their distributed measurement $\mathbb{M}^{\rm AB}$. The free operations for the QRT of Buscemi nonlocality are the so-called Local Operations and Shared Randomness (LOSR) \cite{DS1, DS2, DS3, DS4}. There, Alice and Bob are allowed to share any amount of classical memory described by a random variable $\lambda$. Formally this is specified by providing a probability distribution $p(\lambda)$ which is available to both parties. Moreover, before measuring their systems both parties are allowed to locally perform any completely positive and trace-preserving map, potentially conditioned on the value of the shared memory, i.e. we allow for applying $\mathcal{E}_{\lambda}$ on Alice's and $\mathcal{N}_{\lambda}$ on Bob's side. Finally, the parties are allowed to post-process their measurement outcomes using arbitrary classical channels $p(a|i, \lambda)$ and $p(b|j, \lambda)$ to produce their final guesses. This procedure leads to the most general type of LOSR operation that can be performed on a measurement of the form (\ref{eq:quantum_meas}) \cite{DS3,DS4}. In what follows we will refer to this as \emph{quantum simulation}:
\begin{definition}
\label{def_quantum_sim} (Quantum simulation)
 A quantum simulation of a bipartite measurement $\mathbb{M} = \{M_{ij}\}$ with a subroutine:
 \begin{align}
 \label{eq:quantum_sim}
 \mathscr{S} = \{p(\lambda), p(a|i, \lambda), p(b|j, \lambda), \mathcal{E}_{\lambda}, \mathcal{N}_{\lambda}\}   
 \end{align}
 is a transformation which maps the POVM elements of $\mathbb{M}$ into:
 \begin{align}
    \label{def:quantum-sim2}
     M_{ab}' = \sum_{i,j,\lambda} p(\lambda)p(a|i, \lambda) p(b|j, \lambda)  (\mathcal{E}^{\dagger}_{\lambda} \ot \mathcal{N}^{\dagger}_{\lambda}) [M_{ij}],
 \end{align}
 where $\mathcal{E}^{\dagger}$ denotes the (unique) dual map to $\mathcal{E}$.
\end{definition}
\noindent In other words, any action that can be performed by Alice and Bob in their labs without access to communication can be described by some quantum simulation subroutine. 

Quantum simulation induces a natural preorder on the set of all bipartite measurements. Formally, a preorder is an ordering relation that is reflexive $(a \succ a)$ and transitive $(a \succ b)$ and $(b \succ c)$ implies $(a \succ c)$. Here the preorder induced by quantum simulation will be denoted with $\succ_{\text{q}}$, i.e. $\mathbb{M} \succ_{\text{q}} \mathbb{M}' $ if and only if there there exists a subroutine $\mathscr{S}$ which allows $\mathbb{M}$ to simulate $\mathbb{M}'$, i.e. for the two measurements $\mathbb{M}$ and $\mathbb{M}'$, condition (\ref{def:quantum-sim2}) in Definition \ref{def_quantum_sim} holds. The notion of simulation will turn out to be relevant for the operational tasks introduced later on.


\subsection{Nonlocality from the perspective of no-signalling games}
\label{non-local-games}
Bell nonlocality can be best understood from the perspective of no-signalling games, which also provides an intuitive understanding of Bell inequalities. Such games have been extensively studied in computer science for a long time, where they are a special instance of \emph{interactive proof systems} \cite{Cleve2004}.

The standard scenario of a no-signalling game involves two cooperating players (Alice and Bob) who play the game against a third party, the referee. The referee chooses a question $x \in \mathcal{X}$ for Alice and $y \in \mathcal{Y}$ for Bob according to some probability distribution $p(x,y): \mathcal{X} \times \mathcal{Y} \rightarrow [0, 1]$, where $\mathcal{X}$ and $\mathcal{Y}$ denote finite sets of questions.  Without communicating, and therefore, without knowing what question the other player was asked, Alice (Bob) returns an answer $a \in \mathcal{A}$ $(b \in \mathcal{B})$ from a finite set of possible answers $\mathcal{A}$ $(\mathcal{B})$. Based on the questions asked and the received answers, the referee determines whether the players win or lose the game, according to a pre-arranged set of rules. Such rules are typically expressed using a function $V:$ $\mathcal{A} \times \mathcal{B} \times \mathcal{X}\times \mathcal{Y} \rightarrow [0, 1]$, where $V(a,b,x,y) = 1$ if and only if Alice and Bob win the game by answering $a$ and $b$ for questions $x$ and $y$. 

Alice and Bob know the rules of the game, that is, they know the function $V$ and the distribution of questions $p(x,y)$. Before the game starts they can agree on any strategy which provides them with the best chances of winning.  However, once the game starts, they are not allowed to communicate any more. In the classical setting any strategy they can possibly devise can be encoded in a classical memory system, represented by a shared random variable $\lambda$ and a probability distribution $p(\lambda)$. In the more general quantum case, any possible strategy can be described by a shared quantum state $\rho$ and a choice of local measurements. 

In order to relate the above game setting with Bell inequalities note that the referee's questions $x$ and $y$ can be thought of as labels for different measurement settings. Similarly, the answers correspond to the outcomes of local measurements. Any measurement strategy (be it classical or quantum) leads to a conditional probability $p(a, b|x, y)$ which describes when Alice and Bob give answers $a$ and $b$ for questions $x$ and $y$, respectively. In the language of Bell inequalities $p(a, b|x,y)$ determine the probability that Alice and Bob obtain measurement outcomes $a$ and $b$ when performing the measurements labelled by $x$ and $y$. The average probability that Alice and Bob win, maximized over all possible strategies, can be written as:
\begin{align}
    \label{eq:bell_beh}
    p_{\rm guess}^{V}(\mathcal{G}, \mathbb{M}) = \sum_{a,b,x,y} p(x,y)  p(a,b|x,y) V(a,b,x,y),
\end{align}
where $\mathcal{G} = \{p(x, y), V\}$ defines the game and the conditional probabilities $p(a,b|x,y)$ are related to the local measurements $\{M_{a|x}^{\rm A}\}$ for Alice and $\{M_{b|y}^{\rm B}\}$ for Bob, via the Born rule:
\begin{align}
    p(a,b|x,y) = \tr[\left(M_{a|x}^{\rm A'} \ot M_{b|y}^{\rm B'}\right)\rho^{\rm A'B'}].
\end{align}
With this in mind, Bell inequalities can be thought of as upper bounds on the average guessing probability $p_{\rm guess}(\mathcal{G}, \mathbb{M})$ with which Alice and Bob can win a nonlocal game $\mathcal{G}$ using a classical strategy (i.e. when $\rho^{\rm A'B'}$ is a separable state), optimized over all local measurements $\{M_{a|x}^{\rm A'}\}$ and $\{M_{b|y}^{\rm B'}\}$ . A violation of a Bell inequality corresponds to the situation when there is a quantum strategy which uses an entangled shared state and outperforms the best classical strategy in a particular game $\mathcal{G}$. 

Importantly, there are entangled states which can never violate any Bell inequality \cite{Werner89, Barrett2002, review_EL}. In the language of no-signalling games this means that there are states $\rho^{\rm A'B'}$ which, although entangled, can never outperform the best classical strategy. However, in \cite{Buscemi2012} Buscemi showed that when we modify the rules of the no-signalling game and allow the referee to ask \emph{quantum} instead of classical questions, then all entangled states can outperform the best classical strategy in some nonlocal game, or equivalently, violate the corresponding Bell inequality. 

Before going into the details, let us note that ``asking classical questions'' can also be mathematically modelled by sending states from a collection of orthogonal states from a fixed basis, e.g. $\{\ket{x}\}$ such that $\sum_x \dyad{x} = \mathbb{1}$ and $\braket{x}{x'} = \delta_{x,x'}$ and similarly for $\{\ket{y}\}$. Such states are perfectly distinguishable and hence Alice and Bob, after receiving their questions, may choose their measurements unambiguously. This can be viewed as giving Alice and Bob the ability to perform controlled bipartite measurements $\mathbb{M}^{\rm AA'} = \{M_{a}^{\rm AA'}\}$ and $\mathbb{M}^{\rm B'B} = \{M_{b}^{\rm B'B}\}$ with the POVM elements:
\begin{align}
    \label{eq:loc_meas_contA}
    & M_{a}^{\rm AA'} = \sum_x \dyad{x}^{A} \ot M_{a|x}^{\rm A'}, \\
    \label{eq:loc_meas_contB}
    & M_{b}^{\rm B'B} = \sum_y M_{b|y}^{\rm B'} \ot \dyad{y}^{B}.
\end{align}
If Alice and Bob share a quantum state $\rho^{\rm A'B'}$ then effectively they have access to a distributed measurement $\mathbb{M}^{\rm AB}$ of the form (\ref{eq:quantum_meas}) and hence their behavior $p(a,b|\omega_x,\omega_y)$ can be written as:
\begin{align}
    p(a,b|\omega_x,\omega_y) 
    &= \tr\left[M_{ab}^{\rm AB} \left(\dyad{x}_{\rm A} \ot \dyad{y}_{\rm B}\right)\right],
    \\
    &= 
    \tr[\left(M_{a|x}^{\rm A'} \ot M_{b|y}^{\rm B'}\right)\rho^{\rm A'B'}],
    \\
    &=
    p(a,b|x,y).
\end{align}
With this in mind we can now formalize the process of asking ``quantum questions''. This happens precisely when the states sent by the referee are chosen from an arbitrary collection of states $\{\omega_x\}$. Crucially, these states need not be distinguishable and so each of them can be in a superposition of different orthogonal states. 

Notice, however, that using quantum states as inputs to the distributed measurement $\mathbb{M}^{\rm AB}$ with local measurements of the form (\ref{eq:loc_meas_contA}) and (\ref{eq:loc_meas_contB}) can only lead to a probabilistic version of the standard no-signalling game, i.e. Alice and Bob randomize their choices of measurements according to the respective overlaps $p(x'|x) = \braket{x'|\omega_x}{x'}$ and $p(y'|y) = \braket{y'|\omega_y}{y'}$. Thus, in order to use the power of asking genuinely quantum questions, one needs to allow for arbitrary bipartite local measurements on both sides. This leads to the general form of a distributed measurement (\ref{eq:quantum_meas}) with the local POVM elements $\{M_a^{\rm AA'}\}$ and $\{M_b^{\rm BB'}\}$ being now fully general bipartite measurements, and therefore a Buscemi behaviour is of the form:
\begin{multline}
    p(a,b|\omega_x, \omega_y) 
    \\
    = 
    \tr[(M_a^{\rm AA'}\ot M_b^{\rm B'B})(\omega_x^{\rm A} \ot  \rho^{\rm A'B'} \ot \omega_y^{\rm B})].
    \label{eq:BN}
\end{multline}
The above extension of a no-signalling game leads to a novel type of nonlocality which was noticed for the first time in \cite{Buscemi2012}. Here we will refer to this type of nonclassical correlations as \emph{Buscemi nonlocality}. In this language the main result of \cite{Buscemi2012} states that all entangled states are Buscemi nonlocal. 

In what follows we present a consistent way of quantifying Buscemi nonlocality. First we define a proxy quantity which quantifies how much Buscemi nonlocality can be evidenced using a fixed distributed measurement. Optimizing this quantity over all choices of local measurements for Alice and Bob gives rise to quantity which measures the maximal degree of Buscemi nonlocality which can be obtained using a given quantum state.

\subsection{Quantitative measure of Buscemi nonlocality}
\label{sec3a}
\noindent The fact that Alice and Bob may share entanglement in (\ref{eq:quantum_meas}) and use it to perform a measurement means that the measurement is inherently nonlocal and can lead to interesting correlations, even when measured on completely independent systems. Our central question then is how to quantify this nonlocality present in a bipartite measurement. To build a valid reference point we first consider the case when the measurement does not lead to any type of quantum correlations. This means that the behavior $p(a,b|\omega_x,\omega_y) = \tr[M_{ab}^{\rm AB}( \omega_x^{\rm A} \ot \omega_y^{\rm B})]$ results from the measurement $\{M_{ab}^{\rm AB}\}$ formed using a separable shared state $\rho^{\rm A'B'} \in \fsep$, where $\fsep$ denotes the set of all separable operators. Any separable state can be written as:
\begin{align}
    \label{eq:classical_m}
    \rho^{\rm A'B'} = \sum_{\lambda} p(\lambda) \, \rho_{\lambda}^{\rm A'} \ot \rho_{\lambda}^{\rm B'},
\end{align}
where $p(\lambda)$ is a classical probability distribution corresponding to a shared random variable $\lambda$ and $\{\rho_{\lambda}^{\rm A'}\}$ and $\{\rho_{\lambda}^{\rm B'}\}$ are collections of local quantum states. The associated distributed measurement from Eq. (\ref{eq:quantum_meas}) takes the form:
\begin{align}
    \label{eq:sep_measurement}
    M_{ab}^{\rm AB} = \sum_{\lambda} p(\lambda)\, M_{a|\lambda}^{\rm A} \ot M_{b|\lambda}^{\rm B},
\end{align}
where we denoted $M_{a|\lambda}^{\rm A} := \tr_{\rm A'}[M_a^{\rm AA'}(\mathbb{1}^{\rm A} \ot \rho_{\lambda}^{\rm A'})]$ for Alice and $M_{b|\lambda}^{\rm B} := \tr_{\rm B'}[M_b^{\rm B'B}(\rho_{\lambda}^{\rm B'} \ot \mathbb{1}^{\rm B})]$ for Bob. This is the most general classical measurement scheme which can be realized if Alice and Bob have access only to classical randomness $\lambda$ and the ability to locally prepare quantum states in their labs. We will denote the set of all such measurements by $\fbn$, in analogy with the set of free objects studied in the context of resource theories. Hence in our notation $\fbn$ is a set of measurements of the form (\ref{eq:sep_measurement}). Notice that measurements from this set have POVM elements that are all separable ($\fsep$) and admit a quantum realization ($\q$), i.e can be written as in (\ref{eq:quantum_meas}) for some choice of local measurements and shared state. Such measurements can never demonstrate Buscemi nonlocality, regardless of the state being measured. 

 A natural question is, given an arbitrary bipartite measurement $\mathbb{M}^{\rm AB} \in \q$, how can its nonlocal properties be quantified, in particular its ability to generate Buscemi nonlocality?  For this purpose it is useful to define the following quantity:
\begin{definition} (Robustness of Buscemi Nonlocality \cite{MDI8})
\label{def2}
 The robustness of Buscemi nonlocality (RoBN) of a distributed measurement $\mathbb{M}^{\rm AB} = \{M_{ab}^{\rm AB}\} $ is the solution to the following optimization problem:
\begin{align}
    \label{eq:primal}
    \robn(\mathbb{M}^{\rm AB}) =  \min  \quad & r \\ \nonumber
     \rm{s.t.} \quad  & M_{ab}^{\rm{AB}} + {r}N_{ab}^{\rm{AB}}  = (1+r)O_{ab}^{\rm{AB}}\quad \forall\, a, b,  \\ \nonumber
    & \{O_{ab}^{\rm{AB}}\} \in \fbn, \quad \{N_{ab}^{\rm{AB}}\}\in \rbn. \nonumber
\end{align}
\end{definition}
\noindent Although this may not seem obvious at first sight, the above is a convex optimization problem and hence can be efficiently solved numerically \cite{convopt_book,Watrous2018,cvx} (see Appendix A for details). Moreover, due to the duality of convex optimization problems the dual formulation of the above has several nice properties which will be useful for our purposes. Robustness-based quantifiers were introduced in \cite{Vidal1999,Steiner_2003} as entanglement quantifiers and since then successfully applied in a wide range of QRTs. The above variant is closely related to the MDI-nonlocality robustness introduced in \cite{MDI8,MDI10} at the level of probabilities (\ref{eq:BN}). In particular, the two quantities are equivalent when the sets of input states $\{\omega_x\}$ and $\{\omega_y\}$ are tomographically-complete. It is also worth mentioning that the quantity defined in Def. \ref{def2} is not a particular case of the robustness defined for general convex resource theories of measurements \cite{Oszmaniec2019operational, Takagi2019}. In particular, in Def. \ref{def2} the optimization is over all measurements $\{N_{ab}^{\rm AB}\}$ and $\{O_{ab}^{\rm AB}\}$ which have a quantum realization in the no-signalling scenario, whereas the quantifiers considered in \cite{Oszmaniec2019operational} allow for arbitrary measurements (in particular also those which require communication). In other words, the above general approach is valid only for measurements performed in a single location, whereas here we are explicitly interested in a distributed, multipartite scenario. Hence our robustness measure is a genuinely different quantity than the generalized robustness of measurements studied in the above papers.

In the Appendix A we derived the dual formulation of the RoBN, which will be used to study its operational characterisation. Furthermore, we note that RoBN  possesses three natural properties which one would expect from a reasonable measure of nonlocality, i.e:
\begin{enumerate}
    \item[($i$)] It is \emph{faithful}, meaning that it vanishes if and only if the measurement is classical, i.e:
    \begin{align}
        \robn(\mathbb{M}^{\rm AB}) = 0 \iff \mathbb{M}^{\rm AB} \in \fbn.
    \end{align}
    \item[($ii$)] It is \emph{convex}, meaning that having access to two distributed measurements $\mathbb{M}_1^{\rm AB}$ and $\mathbb{M}_2^{\rm AB}$ one cannot obtain a better one by using them probabilistically, i.e for $\mathbb{M}^{\rm AB} = p\, \mathbb{M}_1 + (1-p)\,\mathbb{M}_2$ with 0 $\leq p \leq 1$, we have:
    \begin{align}
       \quad \hspace{6pt} \robn\! \left( \mathbb{M}^{\rm AB} \right)\! \leq p\, \robn(\mathbb{\mathbb{M}}_1^{\rm AB})\! +\! (1\!-\!p)\,\robn(\mathbb{M}_2^{\rm AB}).
    \end{align}
    \item[($iii$)] It is \emph{monotonic} (non-increasing) under all quantum simulations. That is, if $\mathbb{N}^{\rm AB}$ can be simulated by $\mathbb{M}^{\rm AB}$ using some quantum simulation strategy (\ref{eq:quantum_sim}) then
    \begin{align}
    \robn(\mathbb{N}^{\rm AB}) \leq \robn(\mathbb{M}^{\rm AB}).
    \end{align}
\end{enumerate}
These properties were proven in \cite{DS4} for a more general class of objects. For completeness, we given an independent proof in Appendix B. 

Finally, we introduce a quantity which measures how much Buscemi nonlocality can be generated by using a fixed state. In this way we define the \emph{robustness of Buscemi nonlocality} of a state $\rho_{\rm AB}$ as:
\begin{align}
    \robn(\rho_{\rm AB}) = \max_{\mathbb{M}^{\rm A}, \,\, \mathbb{M}^{\rm B}} \robn(\mathbb{M}^{\rm AB}),
\end{align}
where the optimization ranges over all local measurements on Alice's and Bob's side, $\mathbb{M}^{\rm AB}$ is a distributed measurement of the form (\ref{eq:quantum_meas}) and $\robn(\mathbb{M}^{\rm AB})$ is the robustness quantifier defined in (\ref{eq:primal}). 

\section{Results}
\subsection{Operational characterisation of RoBN}
\label{sec3b}
In the previous section we introduced a measure of Buscemi nonlocality quantifying how ``close'' a given measurement is to that which would arise from using only local measurements and shared randomness, i.e. a measurement of the form (\ref{eq:sep_measurement}). In what follows we will show that RoBN quantifies the advantage offered by a fixed distributed measurement over all classical measurements in a special type of a state discrimination task relevant in the distributed scenario. 

Let us now consider a task which is a special case of the no-signalling game described in Sec. \ref{non-local-games}. In this case we choose the function $V(a,b,x,y) = \delta_{ax} \, \delta_{by}$. This means that Alice and Bob win if they both manage to guess the values of $x \in \mathcal{X}$ and $y \in \mathcal{Y}$ which were supplied to them by the referee. This is a variation of the standard state discrimination task in which a single player has to guess the realization of a single random variable $x$. Interestingly, due to the assumption that the players cannot communicate, distributed state discrimination cannot be reduced to the standard state discrimination task. 

\begin{task}(Distributed state discrimination (DSD))  The task consists of the following steps:
\begin{enumerate}
    \item The referee chooses a bipartite state from the ensemble $\{p(x,y), \sigma_{xy}\}$ according to $p(x,y)$ and distributes it among parties by sending one part of it to Alice and the other part to Bob.
    \item After receiving their systems, Alice and Bob can preprocess them using arbitrary channels $\{\mathcal{E}_{\lambda}^{\rm A}\}$ and $\{\mathcal{N}_{\lambda}^{\rm B}\}$, potentially conditioned on a shared randomness $\lambda$. 
    \item Alice and Bob apply \emph{fixed} local measurements $\mathbb{M}^{\rm AA'} = \{M_i^{\rm AA'}\}$ and $\mathbb{M}^{\rm B'B} = \{M_j^{\rm AA'}\}$ to their shares of the state $\sigma_{xy}$ and a part of the shared state $\rho^{\rm A'B'}$. They obtain outcomes $i$ and $j$ respectively, which they can postprocess to produce their guesses $a$ and $b$. 
    \item Alice and Bob communicate their guesses $a$ and $b$ to the referee and win the game if they \emph{both} correctly guess, i.e. when $a = x$ and $b = y$. 
    \end{enumerate}
Notice that the second and the third step can be also formulated as allowing Alice and Bob apply any \emph{quantum simulation} (\ref{def:quantum-sim2}) to their distributed measurement $\mathbb{M}^{\rm AB} \in \rbn$. Hence the two players are effectively simulating a distributed measurement, denoted by $\mathbb{N}^{\rm AB} \prec \mathbb{M}^{\rm AB}$. The average probability of discriminating states in this discrimination game as specified by $\mathcal{G} = \{p(x,y), \sigma_{xy}\}$ can be expressed as:
\begin{flalign}
    \label{eq:guessing}
    &p_{\rm guess}^{\rm{DSD}}(\mathcal{G}, \mathbb{M}^{\rm AB}) = &\\ & \hspace{30pt}\max_{\mathbb{N}^{\rm AB} \prec_q \mathbb{M}^{\rm AB}}  \sum_{a,b,x,y} p(x, y) \tr\left[ N_{ab} \sigma_{xy} \right]\! \delta_{xa}\, \delta_{yb} \nonumber,
\end{flalign}
where the optimization ranges over all measurements $\mathbb{N}^{\rm AB}= \{N_{ab}\}$ which can be quantum-simulated using $\mathbb{M}^{\rm AB}$.
\end{task}
Let us now consider two different situations: ($i$) a classical scenario in which the distributed measurement performed by Alice and Bob is classical, i.e. $\mathbb{M}^{\rm AB} \in \fbn$ , and ($ii$)  a quantum scenario in which the measurement performed by Alice and Bob is genuinely quantum, i.e it cannot be written as in (\ref{eq:classical_m}).

In the classical case ($i$) the optimal average probability of guessing which state from the ensemble $\{p(x,y), \sigma_{xy}\}$ was provided can be expressed as:
\begin{align}
    \label{eq:class-guess}
    p^{\rm DSD}_{\rm guess}(\mathcal{G}) = \max_{\mathbb{N}^{\rm AB} \in \fbn} \, p_{\rm guess}^{\rm{DSD}}(\mathcal{G}, \mathbb{N}^{\rm AB}),
\end{align}
Note that the above optimization has to be performed over the convex set of measurements of the form (\ref{eq:sep_measurement}), which is a subset of all separable measurements. 

In the quantum case ($ii$) the above score can be further improved by exploiting Buscemi nonlocality contained in an entangled state which forms the distributed measurement $\mathbb{M}^{\rm AB}$. The maximal amount by which quantum score outperforms classical one can be quantified by studying the ratio:
\begin{align}
    \label{eq:q_adv}
    \max_{\mathcal{G}}\, \frac{ p_{\rm guess}^{\rm{DSD}}(\mathcal{G}, \mathbb{M}^{\rm AB})}{p^{\rm DSD}_{\rm guess}(\mathcal{G})}.
\end{align}
In Appendix C we show that the maximal advantage which Alice and Bob can achieve when using $\mathbb{M}^{\rm AB} \in \rbn$ over the best classical distributed measurement is precisely equal to the robustness of Buscemi nonlocality defined in (\ref{eq:primal}). Formally, we have the following relation:
\begin{result}
\label{res:sd}
Let $\mathbb{M}^{\rm AB} = \{M_{ab}^{\rm AB}\}$ be a distributed measurement and $\mathcal{G} = \{p(x, y), \sigma_{xy}\}$ be an ensemble of bipartite states. Then :
\begin{align}
    \label{eq:res1}
    \max_{\mathcal{G}} \frac{p_{\rm guess}^{\rm{DSD}}(\mathcal{G}, \mathbb{M}^{\rm AB})}{p_{\rm guess}^{\rm DSD}(\mathcal{G})} = 1+\robn(\mathbb{M}^{\rm AB}).
\end{align}
\end{result}
This provides a direct operational meaning for Buscemi nonlocality. The proof of Result \ref{res:sd} consists of three parts. First we use the primal formulation of the problem (\ref{eq:primal}) to show that the advantage from (\ref{eq:res1}) is always upper-bounded by the RoBN. Secondly, we identify a set of properties which characterize all distributed measurements and add them to the optimization problem (\ref{eq:primal}) as superfluous constraints. Finally, using this characterization we obtain a dual formulation of the problem which, after some simplifications, allows us to extract the optimal ensemble of states $\{p(x,y), \sigma_{xy}\}$ which achieves the optimum in (\ref{eq:res1}). The full proof of this result is in Appendix C.

The task of distributed state discrimination is a particular instance of a no-signalling game. In this respect we can further consider an advantage (\ref{eq:q_adv}), with the average score $p_{\rm succ}(\mathcal{G}, \mathbb{M}^{\rm AB})$ given by (\ref{eq:bell_beh}), and optimize it over all ensembles $\mathcal{G}$ \emph{and} scoring functions $V(a,b,x,y)$. This would allow us to find the largest possible advantage which can be achieved in any possible nonsignalling game. In this way Result \ref{res:sd} naturally leads to the following corollary:

\begin{corollary} Let $\mathbb{M}^{\rm AB}$ and $\mathcal{G}$ be defined as above and let $V(a,b,x,y): \mathcal{A} \times \mathcal{B}\times\mathcal{X}\times\mathcal{Y} \rightarrow [0,1]$. Then: 
\begin{align}
    \max_{V, \mathcal{G}} \frac{p_{\rm guess}^{V}(\mathcal{G}, \mathbb{M}^{\rm AB})}{{\displaystyle \, \, \max_{\substack{\mathbb{N}^{\rm AB} \, \emph{s.t.} \\ \rho^{\rm A'B'} \in \fsep}}} p_{\rm guess}^{V}(\mathcal{G}, \mathbb{N}^{\rm AB})} = 1+\robn(\mathbb{M}^{\rm AB}).
\end{align}
\end{corollary}
In this way we can also interpret RoBN as a quantifier of the Buscemi nonlocality contained within a given distributed measurement.

\subsection{Connecting Buscemi nonlocality with other notions of nonclassicality}
\label{sec3c}

In this section we show that Buscemi nonlocality can be viewed as a type of nonlocality which is strictly stronger than two other well-known notions of nonlocal correlations: entanglement and nonclassical teleportation. 

It is worth mentioning that the authors of \cite{DS3} also studied the relationship between Buscemi nonlocality, nonclassical teleportation and entanglement by studying a partial order between objects representing these resources:  distributed measurements for Buscemi nonlocality. teleportation instruments for nonclassical teleportation and bipartite states for entanglement. Here we address an analogous problem using a more direct approach: we relate robustness quantifiers of these resource theories and find a direct and simple relationship between them.

Recall that a distributed measurement is composed of two local bipartite measurements and a shared state. This setting is very similar to the teleportation protocol in which Alice locally measures an input state provided by the referee and a part of an entangled state which she shares with Bob. Since the resource used in the teleportation task is effectively ``contained'' in the resource which is used in the task of distributed state discrimination, it is natural to ask if we can see some connection between these two tasks. In particular, how is the ability of performing nonclassical teleportation related to the ability of demonstrating Buscemi nonlocality? 
Furthermore, since teleportation is intrinsically related with entanglement \cite{Cavalcanti2017}, also Buscemi nonlocality should be quantitatively related to the entanglement content of a state. In the next section we will show that in fact these three notions of nonclassical correlations are inherently connected and all describe different types of nonlocality. 

\subsubsection{Buscemi nonlocality and nonclassical teleportation}
Quantum teleportation is one of the most important and thought-provoking discoveries in the whole quantum information theory. In the ideal version of the teleportation protocol proposed by Bennett et. al. in \cite{Bennet1993} two players, Alice and Bob, share a
maximally entangled state. A third party, the referee, gives Alice an unknown quantum state. She then performs a Bell-state measurement on that
system and her share of the entangled state and communicates
her measurement result to Bob. With this new information Bob applies an appropriate correcting unitary to his share of the entangled state, transforming it into the state which was initially given to Alice. This protocol can be naturally generalized to more realistic scenarios in which the shared entangled state and measurements performed by Alice are arbitrary.

Teleportation experiment can be also viewed as a way of testing nonlocality of a pair of objects: a state and measurement. In particular, the ``teleportation resource'' in that case is the teleportation channel or, more precisely, a collection of subchannels which form a \emph{teleportation instrument} constructed using the shared state and Alice's measurement. Recall that an instrument $\mathbb{E} = \{\mathcal{E}_a\}$ for $\{a = 1, \ldots, o_{\rm A}\}$ is a collection of $o_{\rm A}$ completely positive and trace non-increasing linear maps $\mathcal{E}_a$, so-called subchannels, such that $\sum_{a=1}^{o_{\rm A}} \mathcal{E}_a$ is a
channel. It was recently shown that the nonlocality present in a teleportation instrument can be exploited in several quantum-information theoretic tasks \cite{PhysRevResearch.2.023029}. In order to relate nonclassical teleportation with Buscemi nonlocality we first introduce the notion of a teleportation instrument.

\begin{definition} (Teleportation instrument)
 A teleportation instrument $\mathbb{\Lambda}^{\rm A \rightarrow B'}$ from Alice to Bob is a collection of subchannels $\{\Lambda_i^{\rm A \rightarrow B'}\}$ defined as:
 \begin{align}
     \label{eq:telep-instr}
    \!\Lambda_i^{\rm A \rightarrow B'}[\omega^{\rm A}] = \tr_{{\rm AA'}} \!\left[(M_a^{\rm{AA'}} \ot \mathbb{1}^{\rm{B'}}) \left(\omega^{\rm{A}} \ot \rho^{\rm{A'B'}} \right)\right].
 \end{align}
\end{definition}

The above notion fully captures the type of channel obtained during the generalized teleportation experiment. For some applications it may be easier to work with states rather than subchannels. In that case for a collection of input states $\{\omega_x^{\rm A}\}$ one can consider the so-called teleportation assemblages (teleportages) $\{\tau_{a|x}^{\rm B'}\}$, where the elements of the assemblage are given by $\tau_{a|x}^{\rm B'} := \Lambda_a^{\rm A \rightarrow B'}[\omega_x^{\rm A}]$.  

Notice that any teleportation instrument satisfies its own 'no-signalling' constraint, which now reads:  $\sum_{i}{\Lambda_i}^{\rm A \rightarrow B'}[\omega^{\rm A}] = \tr_{\rm A'}[\rho^{\rm A'B'}]$ for all input states $\omega^{\rm A}$. In fact, it can also be shown that teleportation instruments are the most general type of no-signalling instruments acting between two parties \cite{PhysRevResearch.2.023029}. A teleportation instrument $\Lam{}^{\rm A \rightarrow \rm B'}$ is said to be classical (or free) if it describes a teleportation experiment performed using a separable shared state. We can find a general form of a classical teleportation instrument by taking $\rho^{\rm A'B'} = \sum_{\lambda} p_{\lambda} \, \rho_{\lambda}^{\rm A'} \ot \rho_{\lambda}^{\rm{B'}}$. The associated (classical) teleportation instrument reads: 
\begin{align}
\Lambda^c_a (\omega_x) &= \nonumber \sum_{\lambda} p_{\lambda} \tr_{\text{VA}} \left[ \left(M_a^{\text{VA}} \ot \mathbb{1}^{\text{B}}\right) \left(\omega_x \ot \rho_{\lambda}^{\text{A}} \ot \rho_{\lambda}^{\text{B}} \right) \right] \\ \label{eq:16}
&= \sum_{\lambda} p_{\lambda}\, p(a|x, \lambda) \, \rho_{\lambda}^{\text{B}},
\end{align}
where $p(a|x, \lambda) = \tr[M_a^{\text{VA}} (\omega_x^{\text{V}} \ot \rho_{\lambda}^{\text{A}})]$. This is the most general classical teleportation scheme which can be realized if Alice and Bob have access only to classical randomness $\lambda$ and the ability to locally prepare quantum states in their labs. In what follows we will denote the set of all instruments which can be written as in (\ref{eq:16}) by $\ft$. If a teleportation instrument cannot be written in this way, we will refer to it as ``nonclassical'' and denote the set of all such instruments with $\rt$. The quantity which quantitatively measures the amount of nonclassicality associated with a given teleportation instrument is called Robustness of Teleportation (RoT) \cite{Cavalcanti2017}. For a teleportation instrument $\Lam{}^{\rm A \rightarrow B'} = \{\Lambda_a^{\rm A \rightarrow B'}\}$ it is defined as: 
\begin{align}
\label{eq:rob_tel}
    \rot(\Lam{}^{\rm A \rightarrow B'}) = \\ \nonumber \min_{r, \,\{\Gamma_a^{\rm A \rightarrow B'}\}, \{\Omega_a^{\rm A \rightarrow B'}\}} \, &r \\ \nonumber \text{s.t.} \hspace{30pt} &\Lambda_a^{\rm A \rightarrow B'} \!\!+ r\, \Omega_a^{\rm A \rightarrow B'} \!\!= (1+r) \, \Gamma_a^{\rm A \rightarrow B'} \,\, \forall\, a, \\ \nonumber & \{\Gamma_a^{\rm A \rightarrow B'}\} \in \ft, \quad \{\Omega_a^{\rm A \rightarrow B'}\} \in \rt. 
\end{align}
It turns out that the above is also a convex optimization problem which can be seen by formulating the constraints using the Choi-Jamiołkowski isomorphism (see Appendix D for details). With the above notation we can now address our next result which relates Buscemi nonlocality with nonclassical teleportation.
\begin{result}
\label{res:robn_rot}
Let $\mathbb{M}^{\rm AB}$ be a distributed measurement composed of local bipartite measurements $\mathbb{M}^{\rm A}$ and $\mathbb{M}^{\rm B}$ and a shared state $\rho^{\rm A'B'}$. Then:
\begin{align}
    \label{eq:rbn_rt}
    \max_{\mathbb{M}^{\rm B}} \quad \robn(\mathbb{M}^{\rm AB}) &= \rot(\mathbb{\Lambda}^{\rm A \rightarrow B'}),
\end{align}
where the optimization is over all local measurements $\mathbb{M}^{\rm B} = \{M_{b}^{\rm B'B}\}$ for Bob. An analogous result holds for a teleportation instrument $\Lam^{\rm B \rightarrow A'}$ if we instead optimize the LHS of Eq. (\ref{eq:rbn_rt}) over all local measurements for Alice. 
\end{result}

The proof of this result is in Appendix D. Let us now use this result to show a new operational interpretation of the above teleportation quantifier.

Consider a task involving two players, Alice and Bob, who have access to a teleportation instrument $\mathbb{\Lambda}^{\rm A \rightarrow B'}$ connecting their labs. Let the referee be in possession of an ensemble of bipartite quantum states $\mathcal{G} = \{p(x,y), \sigma_{xy}\}$. Just as before, the players may discuss on their strategy before the game begins. This means that they may use a shared classical memory $\lambda$ with a corresponding distribution $p(\lambda)$ and conditioning on it Alice may apply one of the channels $\{\mathcal{E}_{\lambda}^{\rm A}\}$ to the input of the teleportation instrument and Bob may apply $\{\mathcal{N}_{\lambda}^{\rm B'}\}$ to the output. The crucial difference here between the standard teleportation protocol is that Bob \emph{does not} know Alice's measurement outcome and so his correction cannot depend on it. The task posed between Alice and Bob is the following: 

\begin{task}(Teleportation-assisted state discrimination (TSD)) The task consists of the following steps:
\begin{enumerate}
    \item The referee chooses a bipartite state from the ensemble $\mathcal{G} = \{p(x,y), \sigma_{xy}\}$ according to $p(x,y)$ and distributes it among parties by sending one part of it to Alice and the other part to Bob.
    \item Alice sends her part of the state to Bob using a teleportation instrument $\mathbb{\Lambda}^{\rm A \rightarrow B'}$.  She is also allowed to pre-process her part of the state conditioned on the classical randomness $\lambda$ using a collection of channels $\{\mathcal{N}^{\rm A}_{\lambda}\}$. Based on the outcome of the teleportation instrument $i$ and potentially $\lambda$ she produces a guess $a$ via $p(a|i, \lambda)$.
    \item Bob applies a correction $\{\mathcal{E}_{\lambda}^{\rm B'}\}$ conditioned on the value of a shared random variable $\lambda$ to the teleported state he received from Alice. He then measures both parts of the system using an \emph{arbitrary} measurement $\mathbb{M}^{\rm B} = \{M_b^{\rm BB'}\}$ and produces a guess $b$.
    \item Alice and Bob win the game if they both simultaneously guess $x$ and $y$.
\end{enumerate}
\noindent The average probability of guessing in the above discrimination task can be expressed as:
\begin{align}
    &p_{\rm guess}^{\rm{TSD}}(\mathcal{G}, \mathbb{{\Lambda}}^{\rm A \rightarrow B'})= \\ &\max_{\mathbb{M}^{\rm B}} \max_{\mathbb{\Phi} \prec_q \mathbb{\Lambda}} \sum_{a,b,x,y}\!\! p(x, y)  \tr\!\left[ M_b^{\rm B'B} (\Phi_a^{\rm A \rightarrow B'} \!\!\ot \id^{\rm B})\, \sigma_{xy}^{\rm AB} \right] \delta_{xa} \delta_{yb} \!\! \nonumber
\end{align}
where the optimization ranges over all measurements $\mathbb{M}^{\rm B} = \{M_{b}^{\rm B'B}\}$ on Bob's side and all teleportation instruments $\mathbb{\Phi}^{\rm A \rightarrow B'} = \{{\Phi}_a^{\rm A \rightarrow B'}\}$ which can be quantum-simulated using the instrument $\mathbb{\Lambda}^{\rm A \rightarrow B'} = \{{\Lambda}_i^{\rm A \rightarrow B'}\}$. The elements of such a simulated instrument are of the form:
\begin{align}
    {\Phi}_a^{\rm A \rightarrow B'}[\cdot] = \sum_{i, \lambda} p(\lambda) p(a|i, \lambda) \circ \mathcal{N}_{\lambda}^{\rm A} \circ \Lambda_i^{\rm A \rightarrow B'} \circ \mathcal{E}_{\lambda}^{\rm B'}[\cdot]
\end{align}
for some choice of local channels $\{\mathcal{E}_{\lambda}^{\rm B'}\}$, $\{\mathcal{N}_{\lambda}^{\rm A}\}$ and probabilities $p(a|i, \lambda)$ and $p(\lambda)$.
\end{task}

The optimal average probability of guessing that can be achieved using only classical resources (i.e. a separable shared state, meaning that the teleportation instrument is classical) can be written as:
\begin{align}
    p_{\rm guess}^{\rm TSD}(\mathcal{G}) = \max_{\mathbb{F}^{\rm A \rightarrow B'} \in \ft} p_{\rm guess}^{\rm TSD}(\mathcal{G}, \mathbb{F}^{\rm A \rightarrow B'}),
\end{align}
where $\mathbb{F}^{\rm A \rightarrow B'}$ stands for a classical teleportation instrument from Alice to Bob. The maximal advantage which can be offered by any resourceful teleportation instrument $\mathbb{\Lambda}^{\rm A \rightarrow B'}$ in the task of TSD is precisely equal to the quantifier of nonclassical teleportation defined in (\ref{eq:rob_tel}). This is captured by the following result:
\begin{result}
\label{res:td}
Let $\mathbb{\Lambda}^{\rm A \rightarrow B'} = \{\Lambda_{a}^{\rm A \rightarrow B'}\}$ be a teleportation instrument from Alice to Bob and let $\mathcal{G} = \{p(x,y), \sigma_{xy}\}$ be an ensemble of bipartite states. Then the following holds:
\begin{align}
    \max_{\mathcal{G}} \, \frac{p_{\rm guess}^{\rm{TSD}}(\mathcal{G}, \mathbb{\Lambda^{\rm A \rightarrow B}})}{p_{\rm guess}^{\rm TSD}(\mathcal{G})} = 1+\rot(\mathbb{\Lambda}^{\rm A \rightarrow B}).
\end{align}
\end{result}
\begin{proof}
Consider maximizing both sides of Eq. (\ref{eq:res1}) over all measurements $\mathbb{M}^{\rm B}$ on Bob's side. Due to the Result \ref{res:robn_rot}, the right-hand side of Eq. (\ref{eq:res1}) is equal to $1+\rot(\Lam{}^{\rm A\rightarrow B'})$. On the other hand, notice that we can interchange maximisation over $\mathcal{G}$ with maximisation over $\mathbb{M}^{\rm B}$. Since $p_{\rm guess}^{\rm DSD}(\mathcal{G})$ does not depend on $\mathbb{M}^{\rm B}$, the left-hand side of Eq. (\ref{eq:res1}) becomes: 
\begin{align}
    \max_{\mathcal{G}} \frac{\displaystyle\max_{ \mathbb{M}^{\rm B}} p_{\rm guess}^{\rm DSD}(\mathcal{G}, \mathbb{M}^{\rm AB})}{p_{\rm guess}^{\rm DSD}(\mathcal{G})} &=  \max_{\mathcal{G}} \frac{p_{\rm guess}^{\rm TSD}(\mathcal{G}, \mathbb{\Lambda}^{\rm A\rightarrow B'})}{p_{\rm guess}^{\rm DSD}(\mathcal{G})}
    \\ &= \max_{\mathcal{G}} \frac{p_{\rm guess}^{\rm TSD}(\mathcal{G}, \mathbb{\Lambda}^{\rm A\rightarrow B'})}{p_{\rm guess}^{\rm TSD}(\mathcal{G})},
\end{align}
where the last equality follows since:
\begin{align}
    \label{eq:dsd_tsd}
    p^{\rm DSD}_{\rm guess}(\mathcal{G}) &= \max_{\mathbb{F}^{\rm AB} \in \mathcal{F}_{\rm BN}} \, p_{\rm guess}^{\rm{DSD}}(\mathcal{G}, \mathbb{F}^{\rm AB}) \\
    &= \max_{\mathbb{F}^{\rm A\rightarrow B'} \in \mathcal{F}_{\rm T}} \max_{\mathbb{M}^{\rm B}} p_{\rm guess}^{\rm{TSD}}(\mathcal{G}, \mathbb{F}^{\rm A \rightarrow B'})\\
    &= p_{\rm guess}^{\rm{TSD}}(\mathcal{G}).
\end{align}
This completes the proof.
\end{proof}
\subsubsection{Buscemi nonlocality and entanglement}
Let us now explore the link between Buscemi nonlocality, which we defined as a property of a bipartite state and local measurements, and entanglement (a property of the state only). Among the large variety of known entanglement quantifiers \cite{plenio2005introduction,Horodecki_2009,Bennett_1996,Peres1996,Vedral_1997,Rains1999,schumacher2000relative,Hayden_2001}, we are going to choose the one which most naturally relates to the RoBN --- the so-called generalized Robustness of Entanglement (RoE), denoted here with $\roe(\rho)$. This entanglement quantifier was considered for the first time in \cite{Vidal1999} and generalized in \cite{Steiner_2003} and since then proved to be useful in several different contexts, e.g. in proving that all entangled states can demonstrate nonclassical teleportation \cite{Cavalcanti2016}, in exploring the connection between entanglement and permutation symmetry \cite{PhysRevA.65.032328} or in studying the effects of local decoherence on multi-party entanglement \cite{PhysRevA.65.052327}. This quantifier also has two interesting operational interpretations: it quantifies the maximal advantage that can be achieved in a bipartite subchannel discrimination task \cite{PhysRevLett.122.140402} and the maximal advantage in the task of local subchannel discrimination with a quantum memory \cite{PhysRevResearch.2.023029}. It is defined in terms of the following convex optimization problem:
\begin{align}
\label{eq:rob_ent}
    \mathcal{R}_{\rm E}(\rho^{\rm AB}) = \min_{r, \eta^{\rm AB}, \sigma^{\rm AB}} \quad &r \\ \nonumber \text{s.t.} \qquad & \rho^{\rm AB} + r\, \eta^{\rm AB} = (1+r) \sigma^{\rm AB} \\ \nonumber & \eta^{\rm AB} \geq 0, \quad \tr \eta^{\rm AB}=1 \\ 
    &\sigma^{\rm AB} \in \fsep, \quad \tr \sigma^{\rm AB} = 1. \nonumber 
\end{align}
Using this definition we can now address our next result which relates Buscemi nonlocality with entanglement.
\begin{result}
\label{res:robn_roe}
Let $\mathbb{M}^{\rm AB}$ be a distributed measurement composed of local measurements $\mathbb{M}^{\rm A}$ and $\mathbb{M}^{\rm B}$ and a shared state $\rho^{\rm A'B'}$. Then:
\begin{align}
    \max_{\mathbb{M}^{\rm A}, \mathbb{M}^{\rm B}} \quad \robn(\mathbb{M}^{\rm AB}) &= \roe(\rho^{\rm A'B'}),
\end{align}
where the optimization is over all local measurements for Alice $\mathbb{M}^{\rm A} = \{M_{a}^{\rm AA'}\}$ and for Bob $\mathbb{M}^{\rm B} = \{M_{b}^{\rm B'B}\}$.
\end{result}

The proof of this result is in Appendix E. The above relationship along with Result \ref{res:sd} allows to find a new operational interpretation of the RoE. Consider again the task of DSD with the relaxation that Alice and Bob may now apply arbitrary local measurements in their labs. The goal for Alice and Bob remains the same: to guess which state from the ensemble $\mathcal{G} = \{p(x,y), \sigma_{xy}\}$ was prepared, under the assumption that no communication is allowed. In this way the task posed between Alice and Bob is the following:

\begin{task}(Entanglement-assisted state discrimination (ESD)) The task consists of the following steps:
\begin{enumerate}
    \item The referee chooses a bipartite state from the ensemble $\mathcal{G} = \{p(x,y), \sigma_{xy}\}$ according to $p(x,y)$ and distributes it among parties by sending one part of it to Alice and the other part to Bob.
    \item Alice and Bob apply \emph{arbitrary} local measurements $\mathbb{M}^{\rm A}$ and $\mathbb{M}^{\rm B}$ to the states they received and their part of the shared state $\rho^{\rm A'B'}$ and receive outcomes $a$ and $b$, respectively.
    \item Alice and Bob win the game if they both guess which state was provided, i.e. guess both  $x$ and $y$.
\end{enumerate}
\noindent The average probability of guessing in this task can be expressed as:
\begin{align}
    \label{eq:esd_score}
    p_{\rm guess}^{\rm{ESD}}&(\mathcal{G}, \rho^{\rm A'B'}) = \\  &\max_{\mathbb{M}^{\rm A}, \mathbb{M}^{\rm B}} \sum_{a,b,x,y} p(x, y)  \tr\left[ M_{ab}^{\rm AB} \sigma_{xy}^{\rm AB} \right]  \delta_{xa} \delta_{yb}, \nonumber 
\end{align}
where the optimization ranges over all measurements $\mathbb{M}^{\rm A} = \{M_{a}^{\rm AA'}\}$ on Alice's and $\mathbb{M}^{\rm B} = \{M_{b}^{\rm B'B}\}$ on Bob's side with measurement $M_{ab}^{\rm AB}$ of the form (\ref{eq:quantum_meas}). 
\end{task}

The best average probability of guessing in the classical scenario (i.e. when the shared state is separable) is given by:
\begin{align}
    p_{\rm guess}^{\rm ESD}(\mathcal{G}) &= \max_{\sigma^{\rm A'B'} \in \fsep} p_{\rm guess}^{\rm ESD}(\mathcal{G}, \sigma^{\rm A'B'}) \nonumber \\ 
    &= \max_{\mathbb{N}^{\rm AB} \in \fbn} p_{\rm guess}^{\rm ESD}(\mathcal{G}, \mathbb{N}^{\rm AB}) \nonumber \\
    &= p_{\rm guess}^{\rm DSD}(\mathcal{G}). \label{eq_esd_class} 
\end{align}
The maximal advantage which can be offered by an entangled state $\rho^{\rm A'B'}$ in the ESD task can be quantified using the RoE. This is the content of our next result:
\begin{result}
\label{res:esd}
Let $\rho^{\rm A'B'}$ be a bipartite state shared between Alice and Bob and let $\mathcal{G} = \{p(x,y), \sigma_{xy}\}$ be an ensemble of bipartite states. Then the following holds:
\begin{align}
    \max_{\mathcal{G}} \, \frac{p_{\rm guess}^{\rm{ESD}}(\mathcal{G}, \rho^{\rm A'B'})}{p_{\rm guess}^{\rm ESD}(\mathcal{G})} = 1+\roe(\rho^{\rm A'B'}).
\end{align}
\end{result}
\begin{proof}
The proof of Result \ref{res:esd} proceeds similarly to the case of nonclassical teleportation. Let us maximise both sides of (\ref{eq:res1}) over all measurements on Alice's and Bob's side, i.e. over all $\mathbb{M}^{\rm A}$ and $\mathbb{M}^{\rm B}$. Due to Result \ref{res:robn_roe}, the right-hand side of (\ref{eq:res1}) is equal to $1+ \roe(\rho^{\rm A'B'})$. On the other hand, due to (\ref{eq_esd_class}) we can write the left-hand side of (\ref{eq:res1}) as:
\begin{align}
    \max_{\mathcal{G}} \frac{\displaystyle \max_{\mathbb{M}^{\rm A}, \mathbb{M}^{\rm B}} p_{\rm guess}^{\rm DSD}(\mathcal{G}, \mathbb{M}^{\rm AB})}{p_{\rm guess}^{\rm DSD}(\mathcal{G})} =  \max_{\mathcal{G}} \frac{p_{\rm guess}^{\rm ESD}(\mathcal{G}, \rho^{\rm A'B'})}{p_{\rm guess}^{\rm ESD}(\mathcal{G})}.
\end{align}
This completes the proof.
\end{proof}

\subsubsection{Complete sets of monotones for quantum simulation}
We finish this section by showing that the average guessing probability in the task of DSD completely describes the preorder induced by quantum simulation on distributed measurements $\mathbb{M}^{\rm AB}$. Formally this means that the average guessing probability $p_{\rm guess}^{\rm DSD}(\mathcal{G},\mathbb{M}^{\rm AB})$ when viewed as a function of $\mathcal{G}$ forming a complete set of  monotones for quantum simulation of $\mathbb{M}^{\rm AB}$. This is captured by the following result:
\begin{result}
Any distributed measurement  $\mathbb{M}^{\rm AB}$ can quantum-simulate another measurement ${\mathbb{N}}^{\rm AB}$ if and only if for all ensembles $\mathcal{G} = \{p(x,y), \sigma_{xy}\}$ the following holds:
\begin{align}
\label{res:mon}
    p_{\rm guess}^{\rm DSD}(\mathcal{G},\mathbb{M}^{\rm AB}) \geq p_{\rm guess}^{\rm DSD}(\mathcal{G}, {\mathbb{N}}^{\rm AB}).
\end{align}
\end{result}
\noindent In other words, quantum simulation (or LOSR operations) can never improve the discrimination ability of any distributed measurement. The proof of this result is in Appendix F.

\subsection{RoBN as a quantifier in single-shot information theory}
\label{sec3d}
We now address another way of interpreting RoBN from the point of view of single-shot quantum information theory. In particular, in Appendix G we show that RoBN also quantifies the entanglement-assisted min-accessible information of a quantum-to-classical bipartite channel (i.e. a channel with quantum inputs and classical outputs). This connection parallels analogous results from the literature which correspond to single party quantum-to-classical channels \cite{Skrzypczyk2019,Takagi2019}.  

We start by noticing that any distributed measurement $\mathbb{M}^{\rm AB}$ can be seen as an entanglement-assisted quantum-to-classical channel:
\begin{multline}
    \label{eq:channel_imin}
    \mathcal{N}^{\rm AB \rightarrow XY}[\omega^{\rm A} \ot \omega^{\rm B}] 
    \\
    = 
    \sum_{a,b} 
    p(a,b|\omega_x, \omega_y)
    \dyad{a}^{\rm X} \!\ot \dyad{b}^{\rm Y}\!\!\!, \!
\end{multline}
with $p(a,b|\omega_x,\omega_y)$ as in (\ref{eq:BN}). In quantum information theory the standard quantifier of the maximal amount of classical information that can be reliably sent through a quantum channel is the accessible information which is defined for an arbitrary quantum channel $\mathcal{R}$ as:
\begin{align}
    I^{\rm acc}(\mathcal{R}) = \max_{\mathscr{E}, \mathscr{D}} I(X:G),
\end{align}
where $\mathscr{E} = \{p(x), \sigma_x\}$ is an ensemble of states which encode classical random variable $X$ distributed according to $p(x)$, $\mathscr{D} = \{D_g\}$ is the decoding POVM which produces an outcome $g$ with probability $p(g|x) := \tr[D_g \cdot \mathcal{R}[\sigma_x]]$ and $I(X;G) = H(X) - H(X|G)$ is the mutual information of the distribution $p(x,g) := p(x) p(g|x)$. In the single-shot case a more relevant quantity is the \emph{min-accessible information}  $I^{\rm acc}_{\rm{min} }(\mathcal{R})$ which is defined as \cite{ciganovic2014}:
\begin{align}
    \label{eq:imax}
    I^{\rm {acc}}_{\rm min}(\mathcal{R}) = \max_{\mathscr{E}, \mathscr{D}}\,\, \big[ H_{\text{min}}(X) - H_{\text{min}}(X|G)\big],
\end{align}
where the optimization ranges over the same encodings and decodings as before and single-shot entropies are given by \cite{renner2005security}:
\begin{align}
    H_{\rm{min}}(X) &= -\log \max_x p(x),\\
    H_{\rm{min}}(X|G) &= - \log \left[\sum_{g}\max_{x}p(x,g)\right],
\end{align}

Let us now consider an encoding of a bipartite random variable $X \times Y$, i.e $\mathscr{E} = \{p(x,y), \sigma_{xy}\}$ and the associated decoding $\mathcal{D} = \{D_g\}$ for $g = 1, \ldots, |X|\cdot|Y|$. In Appendix G we show that for this particular setting RoBN quantifies the min-accessible information of the channel $\mathcal{N}^{\rm AB \rightarrow XY}$. Formally, we have the following result: 
\begin{result}
\label{res:imin}
Let $\mathcal{N}^{\rm AB \rightarrow XY}$ be a quantum-to-classical channel of the form (\ref{eq:channel_imin}). Then the following holds:
\begin{align}
     I^{\rm {acc}}_{\rm min}(\mathcal{N}^{\rm AB \rightarrow XY}) = \log[1 + \robn(\mathbb{M}^{\rm AB})]
\end{align}
\end{result}
The proof of this result is in Appendix G. The above result provides an alternative way of interpreting RoBN as the maximal amount of min-mutual information that can be obtained between the input and output of the channel (\ref{eq:channel_imin}) when using it only once.

\section{Conclusions}
\label{sec4}

In this work we studied the notion of Buscemi nonlocality when it is formalized as a quantum resource theory of distributed measurements. We focused on a robustness-based quantifier (RoBN) describing when a given distributed measurement can generate Buscemi-nonlocal correlations. We showed that this quantifier admits a natural operational interpretation. In particular, it quantifies the maximal advantage that a given measurement provides over all classical measurements in the task of distributed state discrimination (DSD). We proved that the average guessing probability in this task fully characterises the partial order induced by quantum simulation on distributed measurements. Moreover, we showed that RoBN can be also viewed as the maximal single-shot capacity offered by an entanglement-assisted measurement channel. 
We also explored the relationship between RoBN and two other quantifiers of nonclassical correlations studied in the literature: the generalized robustnesses of teleportation and entanglement. We proved a direct relationship between RoBN and these two quantifiers. This allowed us to provide new operational interpretations for both of these quantifiers in terms of the appropriately tailored state discrimination tasks of: teleportation-assisted state discrimination (TSD), and entanglement-assisted state discrimination (ESD). 

Finally, we emphasise that while we focused exclusively on quantifying Buscemi  nonlocality using a robustness-based measure, our results can be easily extended to address the so-called \emph{weight-based} resource quantifiers \cite{elitzur1992,Lewenstein1998}. These geometric measures find their operational meaning in the so-called exclusion tasks \cite{DS, uola2019quantum}. Consequently, the resource quantifiers of: weight of Buscemi nonlocality, weight of nonclassical teleportation, and the weight of entanglement, are quantifiers characterising: distributed state exclusion (DSE), teleportation-assisted state exclusion (TSE), and entanglement-assisted state exclusion (ESE), respectively.

We believe that the results presented in this work will shed new light on the complex structure of different types of nonclassical effects observed in Nature, as well as on their practical relevance for physically-motivated tasks. 

This work also provides an example of a multiobject quantum resource theory which cannot be reduced to a theory of either measurements, states, channels, or state-measurement pairs \cite{DLS}. This also means that the composite objects we study here constitute genuine multiobject quantum resources. It is an interesting open question to see if one can find additional examples of multiobject resource theories which address such irreducible resources. This is in sharp contrast to a recently introduced multiobject resource theory of state-measurement pairs, where the resources independently contribute to the benefit of the operational task of discrimination and exclusion of subchannels \cite{DLS}.

One of the standard questions addressed by quantum resource theories is determining when and at what rate a large number of copies of one resource can be converted into another. The fact that multiobject QRTs cannot be seen as resource theories of constituent objects leads a natural question of whether this can be used to improve the existing asymptotic protocols. For example, in the resource theory of nonclassical teleportation one can ask whether $n$ uses of teleportation instrument can lead to a better teleportation than using $n$ copies of the shared state. Similarly we can ask whether access to $n$ uses of a distributed measurement can be in advantageous over using bipartite measurements and $n$ copies of the shared state.  

\begin{acknowledgements}
We thank Denis Rosset and David Schmid for insightful discussions, as well as feedback on the first version of this manuscript. P.L.B. acknowledges support from the UK EPSRC (grant no. EP/R00644X/1). T.P. acknowledges support from the EPSRC (grant no. EP/S139151-108). A.F.D. acknowledges support from COLCIENCIAS 756-2016 and the UK EPSRC (EP/L015730/1). PS acknowledges support from a Royal Society URF (UHQT).
\end{acknowledgements}

\bibliographystyle{apsrev4-2}
\bibliography{refs}

\onecolumngrid
\appendix

\section*{Appendix}

\subsection{Equivalent formulation for the Robustness of Buscemi Nonlocality (RoBN)}
By definition RoBN is a conic program. This means that we can use the tools of convex optimization theory to find its dual and from that obtain useful information about the primal problem. We will assume a knowledge of the tools of conic programming, and direct the interested reader to \cite{Boyd2004}. Let us start from the formulation given in the main text and substitute $\widetilde{N}_{ab}^{\rm AB} = r N_{ab}^{\rm AB}$ and $\widetilde{O}_{ab}^{\rm AB} = (1+r) O_{ab}^{\rm AB}$. After this substitution the primal problem can be written as: 
\begin{align}
    \label{eq:primal}
    \robn(\mathbb{M}^{\rm AB}) = \qquad \min  \quad & r \\ \label{eq:App_A_8}
     \rm{s.t.} \quad  & M_{ab}^{\rm{AB}} + \widetilde{N}_{ab}^{\rm{AB}} = \widetilde{O}_{ab}^{\rm{AB}} \qquad \forall \, a, b,  \\ &\{\widetilde{O}_{ab}^{\rm AB}\} \in \fbn, \quad \{\widetilde{N}_{ab}^{\rm AB}\} \in \rbn,
\end{align}
where the optimization is performed over $r, \{\widetilde{N}_{ab}^{\rm{AB}}\}$ and $\{\widetilde{O}_{ab}^{\rm{AB}}\}$. Notice that any collection of operators inside $\rbn$ or $\fbn = \fsep \cap \rbn$ satisfies its own ``no-signalling'' constraint which can be easily deduced from the definition of the set $\rbn$.  Moreover, any operator in $\fbn$ is separable, i.e. $\fbn \in \fsep$. In this way for any $\{{X}_{ab}^{\rm{AB}}\} \in \fbn$ we can write:
\begin{align}
    \label{eq:App_A_1}
    & \sum_{a} X^{\rm AB}_{ab} = \mathbb{1}^{\rm A} \ot X_{b}^{\rm B} \quad \forall\, b \qquad {\rm{and}} \qquad \sum_{b} X_{ab}^{{\rm AB}} = X_{a}^{{\rm A}} \ot \mathbb{1}^{\rm B} \quad \forall \, a \qquad \rm{and} \qquad X_{ab}^{\rm AB} \in \fsep, \\ \label{eq:App_A_2}
    & \sum_b X_b^{\rm B} = \mathbb{1}^{\rm B} \hspace{75pt} {\rm{and}} \qquad \sum_{a} X_a^{\rm A} = \mathbb{1}^{\rm A}.
\end{align}
Now we are going to add a family of such redundant constraints to our optimization problem. Note that we can always do that since adding constraints which are automatically satisfied by any operator in the feasible set does not change the optimal value of the program. 
Moreover, we can also relax the constraint (\ref{eq:App_A_8}) to an inequality $M_{ab}^{\rm AB} + \widetilde{N}_{ab}^{\rm AB} \leq \widetilde{O}^{\rm AB}$ without changing the optimal value of the conic program. To see why this is the case suppose we have solved the relaxed problem using variables $r^{\rm rel}$, $\{\widetilde{N}_{ab}^{\rm AB, rel}\}$, $\{\widetilde{O}_{ab}^{\rm AB, rel}\}$ and $X_{ab}^{\rm AB, rel} \geq 0$ and such that for all $a$ and $b$ we have: $M_{ab}^{\rm AB} + \widetilde{N}_{ab, rel}^{\rm AB} = \widetilde{O}^{\rm AB, rel}_{ab} - X_{ab}^{\rm AB, rel}$. Then the optimal value of the relaxed  program becomes: 
\plb{
\begin{align}
    \robn^{\text{rel}}(\mathbb{M}^{\rm AB}) = -1 +  \frac{1}{d^2} \sum_{ab} \tr \widetilde{O}^{\rm AB,rel}_{a} &= -1 + \frac{1}{d^2} \sum_{ab}\tr \left[M_{ab}^{\rm AB} + \widetilde{N}_{ab}^{\rm AB, rel} + X_{ab}^{\rm AB, rel}\right] \\ 
    &\geq -1 + \frac{1}{d^2} \sum_{ab}\tr \left[M_{ab}^{\rm AB} + \widetilde{N}_{ab}^{\rm AB, rel}\right], \\ 
    &\geq -1 + \frac{1}{d^2} \sum_{ab}\tr \left[M_{ab}^{\rm AB} + \widetilde{N}_{ab}'\right], \\ 
    &\geq -1 + \frac{1}{d^2} \sum_{ab}\tr \left[M_{ab}^{\rm AB} + \widetilde{N}_{ab}^{\rm AB}\right] = \robn(\mathbb{M}^{\rm AB}).
\end{align}
}
where $\{\widetilde{N}_{ab}'\}$ is a set of dual variables feasible for our initial problem \ref{eq:primal}. In this way the conic program defining RoBN becomes:
\begin{align}
    \label{eq:App_A_3}
    \robn(\mathbb{M}) = \qquad \min  \quad & r \\[0.5em] 
    \label{eq:App_A_4}
     \rm{s.t.} \quad  & M_{ab}^{\rm{AB}} + \widetilde{N}_{ab}^{\rm{AB}} \leq \widetilde{O}_{ab}^{\rm{AB}} \hspace{12pt} \forall \, a, b,  \\[0.5em] \label{eq:App_A_5} &\sum_{a} \widetilde{O}_{ab}^{\rm{AB}} = \mathbb{1}^{\rm{A}} \ot \widetilde{O}_b^{\rm{B}} \quad \forall \, b, \qquad
     \sum_b \widetilde{O}_b^{\rm{B}} = (1+r) \mathbb{1}^{\rm{B}}, \\
    & \label{eq:App_A_6} \sum_{b} \widetilde{O}_{ab}^{\rm{AB}} = \widetilde{O}_a^{\rm{A}} \ot \mathbb{1}^{\rm{B}} \quad \forall \, a, \qquad \sum_a \widetilde{O}_a^{\rm{A}} = (1+r) \mathbb{1}^{\rm{A}}, \\ \label{eq:App_A_7}
    &\plb{\{\widetilde{O}_{ab}^{\rm{AB}}\} \in \fbn} \hspace{35pt} \forall \, a,b, \qquad  O_{ab}^{\rm AB} \in \fsep \quad \forall \, a,b, \qquad \{\widetilde{N}_{ab}^{\rm{AB}}\} \in \rbn \quad \forall \, a, b, 
\end{align}
where the minimization is performed over $r, \{\widetilde{O}_{ab}^{\rm{AB}}\}, \{\widetilde{O}_{a}^{\rm{A}}\}, \{\widetilde{O}_{b}^{\rm{B}}\}$ and $\{\widetilde{N}_{ab}^{\rm{AB}}\}$.

In what follows we will denote a dual cone to $\mathscr{R}$ using $\mathscr{R}^*$, that is $\mathscr{R}^* := \{X \, |\, \tr XQ \geq 0$ for all $Q \in \mathscr{R}\}$. We will now write the dual formulation of the above problem. To do so we first write the associated Lagrangian using dual Hermitian variables associated with a corresponding set of constraints: $\{A_{ab}^{\rm{AB}}\}$ such that $A_{ab}^{\rm AB} \geq 0$ for all $a$, $b$,  $\{B_{b}^{\rm{AB}}\}$, $\{C_{a}^{\rm{AB}}\}$,  $D^{\rm{A}} \geq 0$,  $E^{\rm{B}} \geq 0$, \plb{$\{F_{ab}^{\rm{AB}}\} \in \fbn^*$ meaning that $\sum_{ab} \tr[F_{ab}^{\rm AB} X_{ab}^{\rm AB}] \geq 0$ for all $\{X_{ab}^{\rm AB}\} \in \fbn$}, $G_{ab}^{\rm{AB}} \in \fsep^*$ for all $a$, $b$, meaning that $\tr[G_{ab}^{\rm AB} X^{\rm AB}] \geq 0$ for all $a$, $b$ and all separable operators $X^{\rm AB} \in \fsep$ and, finally, $\{H_{ab}^{\rm{AB}}\} \in \rbn^*$. With this the Lagrangian function of the conic program (\ref{eq:App_A_3}---\ref{eq:App_A_7}) becomes: 
\begin{align}
    \mathcal{L} &= r + \sum_{ab} \tr A_{ab}^{\rm{AB}} \left[M_{ab}^{\rm{AB}} + \widetilde{N}_{ab}^{\rm{AB}} - \widetilde{O}_{ab}^{\rm{AB}} \right] + \sum_{b} \tr B_b^{\rm{AB}} \left[\sum_a \widetilde{O}_{ab}^{\rm{AB}} - \mathbb{1}^{\rm{A}} \ot \widetilde{O}_{b}^{\rm{B}} \right] \\ 
    & \quad+ \sum_{a} \tr C_a^{\rm{AB}} \left[\sum_b \widetilde{O}_{ab}^{\rm{AB}} - \widetilde{O}_{a}^{\rm{A}} \ot \mathbb{1}^{\rm{B}} \right] + \tr D^{\rm{A}} \left[\sum_a \widetilde{O}_{a}^{\rm{A}} - (1+r)  \mathbb{1}^{\rm{A}} \right] \\ 
    &\quad + \tr E^{\rm{B}} \left[\sum_b \widetilde{O}_{b}^{\rm{B}} - (1+r)  \mathbb{1}^{\rm{B}} \right] - \sum_{a,b}\tr \left[F^{\rm{AB}}_{ab} \widetilde{O}_{ab}^{\rm{AB}}\right] - \sum_{a,b}\tr \left[G^{\rm{AB}}_{ab} \widetilde{O}_{ab}^{\rm{AB}}\right] - \sum_{a,b}\tr \left[H^{\rm{AB}}_{ab} \widetilde{N}_{ab}^{\rm{AB}}\right] \\
    &= r\cdot\left[1-\tr D^{\rm{A}} - \tr E^{\rm{B}}\right] + \sum_{a,b} \tr \widetilde{N}_{ab}\left[A_{ab}^{\rm{AB}} - H_{ab}^{\rm{AB}}\right] + \sum_{a,b} \tr \widetilde{O}_{ab}\left[-A_{ab}^{\rm{AB}} + B_b^{\rm{AB}} + C_a^{\rm{AB}} - F_{ab}^{\rm{AB}} - G_{ab}^{\rm{AB}}\right] \\
    &\quad + \sum_a \tr O_a^{\rm{A}} \left[D^{\rm{A}}-C_a^{\rm{A}}\right] + \sum_b \tr O_b^{\rm{B}} \left[E^{\rm{B}}-B_b^{\rm{B}}\right] + \sum_{ab} \tr \left[A_{ab}^{\rm{AB}} M_{ab}^{\rm{AB}}\right] - \tr D^{\rm{A}} - \tr E^{\rm B}. 
    \end{align}
By demanding that the terms in the square brackets which appear along with the dual variables vanish we can ensure $\mathcal{L} \leq r$. This leads to the following (dual) conic program:
\begin{align}
    \robn(\mathbb{M}^{\rm AB}) = \qquad \max  \quad &  \sum_{ab} \tr \left[A_{ab}^{\rm{AB}} M_{ab}^{\rm{AB}}\right] - 1 \\
    \rm{s.t.} \quad & C_a^{\rm AB } + B_b^{\rm AB} = A_{ab}^{\rm AB} + F_{ab}^{\rm AB} + G_{ab}^{\rm AB} \quad \forall \, a,b, \\[0.5em] 
    & A_{ab}^{\rm AB} = H_{ab}^{\rm AB} \quad \forall \, a,b, \qquad C_a^{\rm A} = D^{\rm A} \quad \forall \, a, \qquad B_b^{\rm B} = E^{\rm B} \quad \forall \, b, \\[0.5em] 
    & A_{ab}^{\rm AB} \geq 0 \quad \forall \, a, b, \quad \{H_{ab}^{\rm AB}\} \in \rbn^*, \quad \plb{\{F_{ab}^{\rm AB}\} \in \fbn^*}, \quad \tr D^{\rm A} + \tr E^{\rm B} = 1. \nonumber
\end{align}
\plb{Notice now  that the set $\fbn \in \fsep$, which implies that the dual sets satisfy $\fsep^* \in \fbn^*$. Hence without loss of generality we can assume $G_{ab}^{\rm AB} = 0$ for all $a$ and $b$. In this way we can express the above program in the following way:}
\begin{align}
    \label{eq:dual}
    1 + \robn(\mathbb{M}^{\rm AB}) = \qquad \max  \quad &  \sum_{ab} \tr \left[A_{ab}^{\rm{AB}} M_{ab}^{\rm{AB}}\right] \\ 
    \rm{s.t.} \quad & C_a^{\rm AB } + B_b^{\rm AB} - A_{ab}^{\rm AB} = \plb{F_{ab}^{\rm AB} \in \fbn^*}  \quad \forall \, a,b, \\[0.5em]
    & C_a^{\rm A} = D^{\rm A} \quad \forall \, a, \qquad C_{a}^{\rm A}, D^{\rm A} \geq 0 \quad \forall \, a, \\[0.5em]
    & A_{ab}^{\rm AB} \geq 0 \quad\,\,\, \forall \, a, b,  \quad \tr D^{\rm A} + \tr E^{\rm B} = 1. 
\end{align}
Using both primal (\ref{eq:primal}) and dual (\ref{eq:dual}) formulations we can now describe some basic properties of the RoBN.

\subsection{Basic properties of the RoBN}
Here we prove the three basic properties of RoBN highlighted in the main text. 
\paragraph{Faithfulness} If $\mathbb{M}^{\rm AB} \in \fbn$ then we can always choose a feasible $r = 0$ in the primal form (\ref{eq:primal}). Since the solution is always non-negative, $r = 0$ is also optimal.
\paragraph{Convexity} Let $\{N_{ab}^1,\, O_{ab}^1\}$ be optimal primal variables for $\robn(\mathbb{M}^1)$ and similarly let  $\{N_{ab}^2,\, O_{ab}^2\}$ be primal-optimal for $\robn(\mathbb{M}^2)$. Define $\mathbb{M}' = \{M'_{ab} \}$ as a convex combination of the two measurements, that is $M'_{ab} = p \, M_{ab}^1 + (1-p) \, M_{ab}^2$ for each $a$ and $b$. We can construct a set of feasible variables for $\robn(\mathbb{M}')$ in the following way: $N_{ab}' = p \, N_{ab}^1 + (1-p)\, N_{ab}^2$ and $O_{ab}' = p \, O_{ab}^1 + (1-p)\, O_{ab}^2$. Substituting $N_{ab}'$ and $O_{ab}'$ into the constraints of the primal form for $\robn(\mathbb{M})$ shows that this choice is feasible. In this way we obtain an upper bound on $\robn(\mathbb{M}')$: 
\begin{align}
    \robn(\mathbb{M}') \leq \tr \sum_{a,b} N_{ab}' = p\, \cdot \tr \sum_{a,b} N_{ab}^1 + (1-p)\, \cdot \tr \sum_{a,b} N_{ab}^2 = p \cdot \mathcal{R}(\mathbb{M}^1) + (1-p)\cdot \mathcal{R}(\mathbb{M}^2).
\end{align}
\paragraph{Monotonicity} Let us start with the assumption that there is a subroutine:
\begin{align*}
\mathscr{S} = \{p(\lambda), p(a|i, \lambda), p(b|j, \lambda), \mathcal{E}_{\lambda}, \mathcal{N}_{\lambda}\}
\end{align*} 
which allows to simulate $\mathbb{M}'$ using $\mathbb{M}$, i.e. $\mathbb{M} \succ_q \mathbb{M}'$. This means that the POVM elements $\{M_{ab}\}$ of $\mathbb{M}$ can be mapped into:
 \begin{align}
    \nonumber
     M_{ab}' = \sum_{i,j,\lambda} p(\lambda)p(a|i, \lambda) p(b|j, \lambda)  (\mathcal{E}^{\dagger}_{\lambda} \ot \mathcal{N}^{\dagger}_{\lambda}) [M_{ij}]
 \end{align}
Suppose now that we solved the dual problem for $\robn(\mathbb{M}')$ using the optimal dual variables $\{A_{ab}'\}$, $\{B_b'\}$,$\{C_a'\}$, $D'$, $E'$ and $\{F_{ab}'\}$. Using these we construct an educated guess for  $\robn(\mathbb{M})$ in the following way:
\begin{align}
    \label{eq:22}
    &A_{ij}^* = \sum_{a, b, \lambda} p(\lambda) \, p(a|i, \lambda) p(b|j, \lambda)\, \left(\mathcal{E}_{\lambda} \ot \mathcal{N}_{\lambda}\right) [A_{ab}'], 
    \qquad B_j^* = \sum_{b, \lambda} p(\lambda) \, p(b|j, \lambda)\, \left(\mathcal{E}_{\lambda} \ot \mathcal{N}_{\lambda}\right) [B_{b}'], \\
    &C_i^* =  \sum_{a, \lambda} p(\lambda) \, p(a|i, \lambda) \, \left(\mathcal{E}_{\lambda} \ot \mathcal{N}_{\lambda}\right) [C_{a}'], \qquad \qquad \qquad \,\,D^* = \sum_{\lambda} p(\lambda) \,\mathcal{E}_{\lambda} [D'], \\
    & E^* = \sum_{\lambda} p(\lambda)\, \mathcal{N}_{\lambda} [E'], \qquad 
    \qquad \qquad \qquad \qquad \qquad\quad \,\,\, \plb{F_{ij}^* = \sum_{a, b, \lambda} p(\lambda) \, p(a|i, \lambda) p(b|j, \lambda)\, \left(\mathcal{E}_{\lambda} \ot \mathcal{N}_{\lambda}\right) [F_{ab}']}
\end{align}
It can be verified that the above choice of variables is feasible for the dual problem (\ref{eq:dual}). In particular, notice that by construction we have $C_{i}^* + B_j^* - A_{ij}^* = \plb{F_{ij}^*}$ for all $i,j$ since the primed dual variables satisfy the constraints of (\ref{eq:dual}). Furthermore, since $\tr_{\rm B}(\mathcal{E}_{\lambda}\ot\mathcal{N}_{\lambda})[X^{\rm AB}] = \mathcal{E}_{\lambda}[X^{\rm A}]$ we can infer that $\tr_{\rm B}C_i^* = D^*$ and $\tr_{\rm A}B_j^* = E^*$. Moreover, as separable maps preserve both positivity and separability we also have that $A^*_{ij} \geq 0$ for all $i$, $j$  and \plb{$\{F^*_{ij}\} \in \fbn^*$}. Using the proposed set of dual variables we find the following lower bound:
\begin{align}
    1 + \robn(\mathbb{M}) &\geq \sum_{i,j} \tr[M_{ij} A_{ij}^*] \\
    &=  \sum_{a, b, i,j,\lambda} p(\lambda) \, p(a|i, \lambda) p(b|j, \lambda)\, \tr[M_{ij} \cdot \left(\mathcal{E}_{\lambda} \ot \mathcal{N}_{\lambda}\right)[A_{ab}]] \\
    &=  \sum_{a, b, i,j,\lambda} p(\lambda) \, p(a|i, \lambda) p(b|j, \lambda)\, \tr[\left(\mathcal{E}_{\lambda}^{\dagger} \ot \mathcal{N}_{\lambda}^{\dagger}\right)[M_{ij}] \cdot A_{ab}] \\
    &= \sum_{a, b} \tr[M_{ab} A_{ab}] \\
    &= 1 + \robn(\mathbb{M}').
\end{align}
This proves that RoBN is monotonic under quantum simulation.

\subsection{Proof of Result 1}
In this section we prove that RoBN can be seen as a quantifier of the advantage a given distributed measurement provides in the task of distributed state discrimination. To simplify notation in this section we shall omit subsystem labels whenever it is clear from the context. Let us recall that the average guessing probability in the task of distributed state discrimination using a distributed measurement $\mathbb{M}$ can be expressed as:
\begin{align}
    p_{\rm guess}^{\rm{DSD}}(\mathcal{G}, \mathbb{M}) = \,\,\max_{\mathbb{N} \prec_q \mathbb{M}}\,\, \sum_{a,b,x,y} p(x, y) \, \tr\left[ N_{ab} \, \sigma_{xy} \right] \delta_{xa} \delta_{yb},
\end{align}
where the optimization ranges over all measurements $\mathbb{N} = \{N_{ab}\}$ which can be quantum-simulated using $\mathbb{M} = \{M_{ij}\}$, where
\begin{align}
M_{ij} = \tr_{\rm AB} \left[(M_i^{\rm A'A} \ot M_j^{\rm BB'}) (\mathbb{1}^{\rm A'} \ot \rho^{\rm AB} \ot \mathbb{1}^{\rm B'})\right]    
\end{align}
is a distributed measurement and $\mathcal{G} = \{p(x, y), \sigma_{xy}\}$ is an ensemble of bipartite states. Suppose that we have solved the dual problem for RoBN (\ref{eq:dual}) using the set of dual variables $\{A_{ab}\}$, $\{C_{a}\}$, $\{B_{b}\}$, $D$, $E$ and $\{G_{ab}\}$. Notice also that due to the constraints in (\ref{eq:dual}) the matrix $A_{ab}$ is positive semi-definite for all values of $a$ and $b$.  Let us now consider a particular game setting $\mathcal{G}^{*} = \{p^*(x, y),  \sigma^*_{xy}\}$ defined in the following way:
\begin{align}
    C = \sum_{x,y} \tr A_{xy}, \qquad p^*(x, y) = \frac{\tr A_{xy}}{C}, \qquad \sigma_{xy}^* =\frac{A_{xy}}{\tr A_{xy}},
\end{align}
where $x = 1, \ldots, o_{\rm A}$ and $y = 1, \ldots, o_{\rm B}$. The best average guessing probability which can be achieved in the game $\mathcal{G}^*$ using a distributed measurement $\mathbb{M}$ is given by:
\begin{align}
    p_{\rm guess}^{\rm{DSD}}(\mathcal{G}^*, \mathbb{M}) &= \max_{\mathbb{N} \prec_q \mathbb{M}} \sum_{a,b,x,y} p^*(x, y) \, \tr\left[ N_{ab} \, \sigma_{xy}^{*} \right] \delta_{xa} \delta_{yb} \\
    & \geq \sum_{x,y} \frac{\tr A_{xy}}{C}\cdot \tr\left[M_{xy} \, \frac{A_{xy}}{\tr A_{xy}} \right] \\ 
    &=\frac{1}{C}\sum_{x,y}\tr\left[M_{xy} A_{xy} \right] \\
    &= \frac{1}{C} \left[1+\robn(\mathbb{M}^{\rm AB})\right],
    \label{eq:lower_bnd}
\end{align}
where  the inequality in the second line we follows from choosing a particular subroutine $\mathscr{S}$ with $p(\lambda) = 1 / |\lambda|$, $p(a|i, \lambda) = \delta_{a i}$, $p(b|j,\lambda) = \delta_{bj}$ and $\mathcal{E}_{\lambda} = \mathcal{N}_{\lambda} = \rm{id}$.  Let us now look at the corresponding classical (i.e. without access to entanglement) probability of guessing:
\begin{align}
     p_{\rm guess}^{\rm DSD}(\mathcal{G}^*) = \max_{\mathbb{N} \in \fbn}  p^{\rm{DSD}}_{\rm guess}(\mathcal{G}^*, \mathbb{N}) &= \max_{\mathbb{N} \in \fbn} \sum_{x,y} p^*(x,y) \, \tr\left[ N_{xy} \, \sigma_{xy}^* \right] \\
     &=  \frac{1}{C} \max_{\mathbb{N} \in \fbn} \sum_{x,y}  \tr\left[  N_{xy} A_{xy} \right] \\
     &=  \frac{1}{C} \max_{\mathbb{N} \in \fbn} \sum_{x,y}  \tr\left[  N_{xy}(C_x + B_y - \plb{F_{xy}}) \right] \\
     &=  \frac{1}{C} \max_{\mathbb{N} \in \fbn} \left(\sum_{x}  \tr\left[  (N_x \ot \mathbb{1}) C_x\right] + \sum_{y}  \tr\left[  (\mathbb{1} \ot N_y ) B_y\right]
     - \sum_{x,y}  \tr\left[  N_{xy} \plb{F_{xy}}\right]\right) \\
     &\leq \frac{1}{C} \max_{\mathbb{N} \in \fbn} \left(\sum_{x}  \tr\left[N_x D \right] + \sum_{y}  \tr\left[  N_y E\right]\right) \\
     & = \frac{1}{C} \left(\tr D + \tr E\right) \\
     & = \frac{1}{C},
      \label{eq:upper_bnd}
     \end{align}
where the inequality follows since for all $\mathbb{N} \in \fbn$ we have \plb{$\sum_{xy} \tr[N_{xy}F_{xy}] \geq 0$}. Combining bounds (\ref{eq:lower_bnd}) and (\ref{eq:upper_bnd}) leads to:
\begin{align}
    \label{eq:bnd_lower_final}
    \max_{\mathcal{G}} \frac{p_{\rm guess}^{\rm{DSD}}(\mathcal{G}, \mathbb{M})}{p_{\rm class}^{\rm DSD}(\mathcal{G})} \geq \frac{p_{\rm guess}^{\rm{DSD}}(\mathcal{G}^*, \mathbb{M})}{p_{\rm class}^{\rm DSD}(\mathcal{G}^*)} \geq 1 + \robn(\mathbb{M}).
\end{align}
In order to prove the upper bound notice that the first line of constraints in the primal formulation for RoBN (\ref{eq:primal}) implies:
\begin{align}
    \forall\, a, b \qquad M_{ab}' = \widetilde{O}_{ab}' - \widetilde{N}_{ab}',
\end{align}
where $\widetilde{O}_{ab}' = \left[1+\robn(\mathbb{M})\right] O_{ab}'$ for all $a,b$ and $\{O_{ab}'\} \in \fbn$. This allows to write:
\begin{align}
    p_{\rm guess}^{\rm{DSD}}(\mathcal{G}, \mathbb{M}) &= \max_{\mathbb{M}' \prec_q \mathbb{M}} \, \sum_{a,b,x,y} p(x, y) \, \tr\left[ M_{ab}' \, \sigma_{xy} \right] \delta_{xa} \delta_{yb} \\
    &= \max_{\mathbb{M}' \prec_q \mathbb{M}} \sum_{a,b,x,y} p(x, y) \, \tr\left[ (\widetilde{O}_{ab}' - \widetilde{N}_{ab}') \sigma_{xy} \right] \delta_{xa} \delta_{yb} \\
    &\leq \max_{\mathbb{M}' \prec_q \mathbb{M}} \sum_{a,b,x,y} p(x, y) \, \tr\left[ \widetilde{O}_{ab}' \sigma_{xy} \right] \delta_{xa} \delta_{yb} \\
    &= \max_{\mathbb{M}' \prec_q \mathbb{M}} \left[1+\robn(\mathbb{M}')\right] \sum_{a,b,x,y} p(x, y) \, \tr\left[ O_{ab}' \sigma_{xy} \right] \delta_{xa} \delta_{yb} \\
    &\leq \left(\max_{\mathbb{M}' \prec_q \mathbb{M}} \left[1+\robn(\mathbb{M}')\right]\right)\left( \max_{\{O_{ab}\} \in \fbn} \sum_{a,b,x,y} p(x, y) \, \tr\left[ O_{ab} \sigma_{xy} \right] \delta_{xa} \delta_{yb}\right) \\
    &\leq \left[1+\robn(\mathbb{M})\right] p^{\rm DSD}_{\rm guess}(\mathcal{G}),
    \label{eq:bnd_upper_final}
\end{align}
where the last inequality follows from the monotonicity of RoBN under quantum simulation. Combining bounds (\ref{eq:bnd_lower_final}) and (\ref{eq:bnd_upper_final}) yields:
\begin{align}
    \max_{\mathcal{G}} \frac{p_{\rm guess}^{\rm{DSD}}(\mathcal{G}, \mathbb{M})}{p^{\rm DSD}_{\rm guess}(\mathcal{G})} = 1+\robn(\mathbb{M}).
\end{align}

\subsection{Proof of Result 2}
Before proving the result we recall the primal and dual formulation of the RoT quantifier. Let $\mathbb{\Lambda} = \{\Lambda_a\}$ be a teleportation instrument whose elements are defined as:
\begin{align}
    \Lambda_a^{\rm A \rightarrow B'}[\omega] := \tr_{\rm AA'}[(M_a^{\rm A A'} \ot \mathbb{1}^{\rm B})(\omega^{\rm A} \ot \rho^{\rm A'B'})],
\end{align}
for some measurement $M_a^{\rm AA'}$ and a shared state $\rho^{\rm A'B'}$. We denote the set of Choi-Jamiolkowski states corresponding to this of these subchannels with $\{J_a^{\rm VB'}\}$, i.e. each $J_a^{\rm VB'} := (\id^{\rm V}\ot\Lambda_a^{\rm A \rightarrow B'})[\phi_+^{\rm VA}]$ with system $V$ isomorphic to $A$. With these definitions RoT for a teleportation instrument $\mathbb{\Lambda}^{\rm A \rightarrow B'}$ can be written as:
\begin{align}
    \label{rot_def}
\begin{aligned}[t]
    \rot(\Lam{}^{\rm A \rightarrow B'}) = \quad \min  \quad & \tr \widetilde{\sigma}^{\,\rm B'}, \\[0.5em]
     \rm{s.t.} \quad  & J_a^{\rm VB'}  \leq  F_a^{\rm VB'} \quad \forall \, a,  \\
    & \sum_{a} F_a^{\rm VB'} = \frac{\mathbb{1}^{\rm V}}{d} \ot \widetilde{\sigma}^{\,\,\rm B'}, \\
    &F_a^{\rm VB'} \in \fsep \quad \forall a, \quad \widetilde{\sigma}^{\rm B'} \geq 0. \\
\end{aligned}
\Longleftrightarrow\qquad
\begin{aligned}[t]
    \max  \quad & \sum_a \tr\left[A_a^{\rm VB'} J_a^{\rm VB'} \right] - 1,\\
     \rm{s.t.} \quad  & B^{\rm VB'} - A_a^{VB'}  = W_a^{\rm VB'} \in \fsep^* \quad \forall a, \\
    & B^{\rm B'} = \mathbb{1}^{\rm B'}, \quad A_{a}^{\rm VB'} \geq 0 \quad \forall a. \\ 
\end{aligned}
\end{align}
Let us now proceed with the proof of Result 2.
\begin{proof}
As before, the proof consists of two steps. First we will show that $\rot(\Lam{}^{\rm A \rightarrow B'})$ lower bounds $\robn(\mathbb{M}^{\rm AB})$ for a particular choice of local measurement $\mathbb{M}^{\rm B'B}$. Then we will show that for any choice of local measurements on Bob's side $\robn(\mathbb{M}^{\rm AB})$ is never larger than the teleportation quantifier $\rot(\Lam{}^{\rm A \rightarrow B'})$. 

Let $A_a^{\rm VB'} \geq 0$, $W_a^{\rm VB'} \in \fsep^*$ and $B^{\rm VB'}$ be optimal dual variables for $\rot(\Lam{}^{\rm A \rightarrow B'})$. Let $\{U_b^{\rm B}\}$ for $b \in \{1, \ldots, d^2\}$ be a set of Pauli operators with respect to a basis $\{\ket{i}^{\rm B}\}$. Consider the following measurement with $o_{\rm B} = d^2$ outcomes:
\begin{align}
    M_b^{\rm B'B} = (\id^{\rm B'} \ot \mathcal{U}_b^{\rm B})[\phi_+^{\rm B'B}],
\end{align}
where $\mathcal{U}_b[\cdot] := U_{b}(\cdot)U_b^{\dagger}$. We are interested in the lower bound for $\rot(\Lam{}^{\rm A \rightarrow B'})$. Let us choose a set of dual variables in (\ref{eq:dual}) inspired by the optimal dual variables for $\rot(\Lam{}^{\rm V \rightarrow B'})$:
\begin{align}
    &A_{ab}^{\rm AB} = (\id^{\rm A} \ot (\mathcal{U}_b^{\dagger})^{\rm B})[(A_a^{\rm AB})^T], \quad
    \plb{F_{ab}^{\rm AB}} = (\id^{\rm A} \ot (\mathcal{U}_b^{\dagger})^{\rm B})[(W_a^{\rm AB})^T], 
    \quad
    B_b^{\rm AB} = \frac{1}{d}(\id^{\rm A} \ot (\mathcal{U}_b^{\dagger})^{\rm B})[(B^{\rm AB})^T], \\
    &C_a^{\rm AB} = 0,
    \qquad
    D^{\rm B} = \frac{1}{d} \mathbb{1}^{\rm B},
    \qquad
    E^{\rm A} = 0.
\end{align}
It can be verified by direct substitution that the above choice is feasible. \plb{In particular, the above choice for $\{F_{ab}^{\rm AB}\}$ is feasible as $\fsep^* \in \fbn^*$ and both sets are invariant under local unitaries.} This leads to the following chain of inequalities:
\begin{align}
    1 + \max_{\mathbb{M}^{\rm B}} \, \robn(\mathbb{M}^{\rm AB}) &\geq \sum_{ab} \tr[A_{ab}^{\rm AB} M_{ab}^{\rm AB}] \\ 
    & = \sum_{ab}\tr\left[ (\id^{\rm A} \ot (\mathcal{U}_b^{\dagger})^{\rm B})[(A_a^{\rm AB})^T] \cdot \tr_{\rm A'B'}\left[(M_{a}^{\rm AA'} \!\! \ot M_{b}^{\rm B'B})(\mathbb{1}^{\rm A} \ot \rho^{\rm A'B'} \ot \mathbb{1}^{\rm B})\right]\right] \\ 
    & = \sum_{ab}\tr\left[ (\id^{\rm A} \ot (\mathcal{U}_b^{\dagger})^{\rm B})[(A_a^{\rm AB})^T] \cdot \tr_{\rm A'B'}\left[(M_{a}^{\rm AA'} \!\! \ot (\id^{\rm B'} \! \ot \mathcal{U}_b^{\rm B})[\phi_+^{\rm B'B}])(\mathbb{1}^{\rm A} \ot \rho^{\rm A'B'} \ot \mathbb{1}^{\rm B})\right]\right]\\
     & = \sum_{ab}\tr\left[ \mathbb{1}^{\rm A'B'}\ot (\id^{\rm A} \ot (\mathcal{U}_b^{\dagger})^{\rm B})[(A_a^{\rm AB})^T] \cdot \left[(M_{a}^{\rm AA'} \!\! \ot (\id^{\rm B'} \! \ot \mathcal{U}_b^{\rm B})[\phi_+^{\rm B'B}])(\mathbb{1}^{\rm A} \ot \rho^{\rm A'B'} \ot \mathbb{1}^{\rm B})\right]\right]\\
    & = \sum_{ab}\tr \left[\left(\mathbb{1}^{\rm A'B'}\ot (A_a^{\rm AB})^T\right) \left(M_{a}^{\rm AA'} \!\! \ot \phi_+^{\rm B'B}\right)\left(\mathbb{1}^{\rm A} \ot \rho^{\rm A'B'} \ot \mathbb{1}^{\rm B})\right)\right]\\
    & = \frac{1}{d^2}\sum_{ab}\tr \left[A_{a}^{\rm VB'} \cdot \tr_{\rm AA'}\left[(\mathbb{1}^{\rm V} \ot M_{a}^{\rm AA'}\ot \mathbb{1}^{\rm B'})(\phi_+^{\rm VA} \ot \rho^{\rm A'B'})\right]\right]\\  
    & = \sum_a \tr[A_a^{\rm VB'} J_a^{\rm VB'}] \\
    &= 1 + \rot (\Lam{}^{\rm V \rightarrow B'}).
\end{align}
We now prove the upper bound. Notice that for any distributed measurement $\mathbb{M}^{\rm AB}$ we can construct $\mathbb{M}^{\rm VB} := \{M_{ab}^{\rm VB}\}$ such that $M_{ab}^{\rm VB} := d \, \tr_{\rm A}[(\mathbb{1}^{\rm V} \ot M_{ab}^{\rm AB})(\phi_+^{\rm VA} \ot \mathbb{1}^{\rm B})]$. This in turn can be written as:
\begin{align}
    M_{ab}^{\rm{V}\rm{B}} &:= d\, \tr_{\rm A A'B'}\left[\left(\mathbb{1}^{\rm A} \ot M_a^{\rm{A}\rm{A}'}\otimes M_b^{\rm{B}'\rm{B}}\right)\left(\phi_+^{\rm{VA}}\otimes \rho^{\rm A' B'} \otimes \mathbb{1}^{\rm{B}}\right)\right] \\
    &=d\,\tr_{\rm B'}\left[\left(\mathbb{1}^{\rm V} \ot M_{b}^{\rm B'B}\right)\left(J_{a}^{\rm{VB'}}\otimes \mathbb{1}^{\rm{B}}\right)\right].
    \label{eq:App_C_1}
\end{align}
Note that we can always write $J_a^{\rm VB'} \leq [1+\rot(\Lam{}^{\rm A \rightarrow B'})] F_a^{\rm VB'}$, where $\{F_a^{\rm VB'}\}$ are Choi-Jamiolkowski operators of some classical teleportation instrument. This allows us to further rewrite (\ref{eq:App_C_1}) as:
\begin{align}
    M_{ab}^{\rm{V}\rm{B}} &\leq d\, [1+\rot(\Lam{}^{\rm A \rightarrow B'})] \, \tr_{\rm A}[(\mathbb{1}^{\rm V} \ot M_{ab}^{\rm AB})(\phi_+^{\rm VA} \ot \mathbb{1}^{\rm B})] = [1+\rot(\Lam{}^{\rm A \rightarrow B'})]  \, O_{ab}^{\rm VB}.
\end{align}
Where $\{O_{ab}^{\rm VB}\}$ is a free distributed measurement. Hence also $M_{ab}^{\rm AB} \leq [1+\rot(\Lam{}^{\rm A \rightarrow B'})]  \, O_{ab}^{\rm AB}$ for some free distributed measurement $\{O_{ab}^{\rm AB}\}$. This finally allows us to write:
\begin{align}
    \max_{\mathbb{M}^{\rm B}} \, \robn(\mathbb{M}^{\rm AB}) \leq  [1+\rot(\Lam{}^{\rm A \rightarrow B'})] \, \max_{\mathbb{M}^{\rm B}} \, \sum_{ab} \tr[A_{ab}^{\rm AB} O_{ab}^{\rm AB}] \leq [1+\rot(\Lam{}^{\rm A \rightarrow B'})].
\end{align}
This proves the lemma.
\end{proof}

\subsection{Proof of Result 4}
Let us recall that the conic program formulation of RoE is given by:
\begin{align}
    \label{roe_def}
\begin{aligned}[t]
    \roe(\rho^{\rm A'B'}) = \quad \min  \quad & \tr \widetilde{\sigma}^{\,\rm A'B'}, \\[0.5em]
     \rm{s.t.} \quad  & \rho^{\rm A'B'}  \leq  \widetilde{\sigma}^{\, \rm A'B'}  \\
    & \widetilde{\sigma}^{\rm A'B'} \in \fsep. \\
\end{aligned}
\Longleftrightarrow\qquad
\begin{aligned}[t]
    \max  \quad & \sum_a \tr\left[A^{\rm A'B'} \rho^{\rm A'B'} \right] - 1,\\
     \rm{s.t.} \quad  & \mathbb{1}^{\rm A'B'} - A^{\rm A'B'} = W^{\rm A'B'} \in \fsep^*, \\ 
     &A^{\rm A'B'}\geq 0.
\end{aligned}
\end{align}
The proof is based on three parts. First we use Result 2 to connect RoBN with RoT. Then we essentially parallel the steps taken in the proof of Result 2 to link RoT with RoE. It is worth mentioning that the link between RoT and RoE has already been obtained some time ago in \cite{Cavalcanti2017}. Here for convenience we state an independent proof. 

\begin{proof}
Let us begin by noting that Result $2$ implies:
\begin{align}
    \max_{\mathbb{M}^{\rm A}, \mathbb{M}^{\rm B}}\robn(\mathbb{M}^{\rm AB}) = \max_{\mathbb{M}^{\rm A}} \left[\max_{ \mathbb{M}^{\rm B}}\robn(\mathbb{M}^{\rm AB}) \right] = \max_{\mathbb{M}^{\rm A}} \rot (\Lam{}^{\rm A \rightarrow B'})
\end{align}
Let $A^{\rm A'B'} \geq 0$, $W^{\rm A'B'} \in \fsep^*$ be optimal dual variables for $\roe(\rho^{\rm A'B'})$. Let $\{U_a^{\rm A'}\}$ for $a \in \{1, \ldots, d^2\}$ be a set of Pauli operators with respect to a basis $\{\ket{i}^{\rm A'}\}$. Consider the following measurement with $o_{\rm A} = d^2$ outcomes:
\begin{align}
    M_a^{\rm AA'} = (\id^{\rm A} \ot \mathcal{U}_a^{\rm A'})[\phi_+^{\rm AA'}].
\end{align}

We are interested in the lower bound for $\roe(\rho^{\rm A'B'})$, let us construct a set of (potentially sub-optimal) dual variables in the maximization (\ref{roe_def}) using the optimal set of dual variables for $\rot(\Lam{}^{\rm V \rightarrow B'})$, i.e.:
\begin{align}
    &A_{a}^{\rm VB'} = ((\mathcal{U}_a^{\dagger})^{\rm V} \ot \id^{\rm B'})[A^{\rm VB'}], \quad
    W_{a}^{\rm VB'} = ((\mathcal{U}_a^{\dagger})^{\rm V} \ot \id^{\rm B'})[(W^{\rm VB'}], 
    \quad
    B^{\rm VB'} = \frac{1}{d} \mathbb{1}^{\rm VB'}.
\end{align}
It can be verified by direct substitution that the above choice is feasible. This leads to the following chain of inequalities:
\begin{align}
    1 + \max_{\mathbb{M}^{\rm A}} \, \rot(\Lam{}^{\rm A \rightarrow B'}) &\geq \sum_{a} \tr[A_{a}^{\rm VB'} J_{a}^{\rm VB'}] \\ 
    & = \sum_{a}\tr\left[ ((\mathcal{U}_a^{\dagger})^{\rm V} \ot \id^{\rm B'})[A_a^{\rm VB'}] \cdot \tr_{\rm AA'}\left[(\mathbb{1}^{\rm V} \ot M_{a}^{\rm AA'} \!\! \ot \mathbb{1}^{\rm B'})(\phi_+^{\rm VA} \ot \rho^{\rm A'B'})\right]\right] \\ 
    & = \sum_{a}\tr\left[ ((\mathcal{U}_a^{\dagger})^{\rm V} \ot \id^{\rm B'})[A_a^{\rm VB'}] \cdot \tr_{\rm AA'}\left[(\mathbb{1}^{\rm V} \ot (\id^{\rm A} \ot \mathcal{U}_a^{\rm A'} )[\phi_+^{\rm AA'}]  \ot \mathbb{1}^{\rm B'})(\phi_+^{\rm VA} \ot \rho^{\rm A'B'})\right]\right] \\ 
     & =  \sum_{a}\tr\left[ \mathbb{1}^{\rm AA'} \ot ((\mathcal{U}_a^{\dagger})^{\rm V} \ot \id^{\rm B'})[A_a^{\rm VB'}] \cdot (\mathbb{1}^{\rm V} \ot (\id^{\rm A} \ot \mathcal{U}_a^{\rm A'} )[\phi_+^{\rm AA'}]  \ot \mathbb{1}^{\rm B'})(\phi_+^{\rm VA} \ot \rho^{\rm A'B'})\right]\\ 
    & = \frac{1}{d^2} \sum_{a}\tr[((\mathcal{U}_{a}^{\dagger})^{\rm A'} \ot \id^{\rm B'})A^{\rm A'B'}\cdot (\mathcal{U}_{a}^{\rm A'} \ot \id^{\rm B'}) \rho^{\rm A'B'}]\\
    & = \tr[A^{\rm A'B'} \rho^{\rm A'B'}]  \\
    &= 1 + \roe (\rho^{\rm A'B'}).
\end{align}
We now prove the upper bound. Notice that any teleportation instrument $\Lam{}^{\rm A \rightarrow B'}$ expressed using Choi-Jamiolkowski operators $\{J_{a}^{\rm VB'}\}$ satisfies:
\begin{align}
    J_{a}^{\rm VB'} &:= \tr_{\rm VA}\left[\left(M_a^{\rm{V}\rm{A}}\otimes \mathbb{1}^{\rm{B}'} \right) \left(\phi_+^{\rm{A}}\otimes \rho^{\rm A' B'} \otimes \mathbb{1}^{\rm{B}}\right)\right] \\
    &\leq [1 + \roe(\rho^{\rm A'B'})] \tr_{\rm VA}\left[\left(M_a^{\rm{V}\rm{A}}\otimes \mathbb{1}^{\rm{B}'} \right) \left(\phi_+^{\rm{A}}\otimes \sigma^{\rm A' B'} \otimes \mathbb{1}^{\rm{B}}\right)\right] \\
    &= [1 + \roe(\rho^{\rm A'B'})] O_a^{\rm VB'},
    \label{eq:App_C_2}
\end{align}
for some state $\sigma^{\rm A' B'} \in \fsep$ and corresponding (classical) teleportation operators $\{O_{a}^{\rm VB'}\}$. In this way we can write:
\begin{align}
    \max_{\mathbb{M}^{\rm A}} \, [1+\rot(\Lam{}^{\rm A \rightarrow B'})] &= \max_{\mathbb{M}^{\rm A}} \, \max_{\{A^{\rm VB'}_a\}} \sum_{a}\tr[A_{a}^{\rm VB'} J_a^{\rm VB'}] \\
    &\leq [1+\roe(\rho^{\rm A'B'})] \sum_{a} \tr[A_{a}^{\rm VB'} O_{a}^{\rm VB'}] \\
    &\leq [1+\roe(\rho^{\rm A'B'})].
\end{align}
This proves the lemma.
\end{proof}

\subsection{Proof of Result 6}
In this section, unless explicitly specified,  all bipartite operators act on subsystems $\rm A$ and $\rm B$. We begin by assuming that a distributed measurement $\mathbb{M}$ can be used to simulate $\mathbb{M}^*$, that is $\mathbb{M} \succ_q \mathbb{M}^*$. We have:
\begin{align}
    p_{\rm guess}^{\rm{DSD}}(\mathcal{G}, \mathbb{M}) &= \max_{\mathbb{M} \succ_q \mathbb{M}'} \sum_{a,b} p(a, b) \, \tr\left[ M_{ab}' \, \sigma_{ab} \right]  \\
    &\geq \max_{\mathbb{M}^* \succ_q \mathbb{M}'} \, \sum_{a,b} p(a, b) \, \tr\left[ M_{ab}' \, \sigma_{ab} \right]  \\
    &=  p_{\rm guess}^{\rm{DSD}}(\mathcal{G}, \mathbb{M}^*),
\end{align}
since the set $\{\mathbb{M}' |\mathbb{M}^* \succ_q  \mathbb{M}' \}$ is a subset of $\{\mathbb{M}' | \mathbb{M} \succ_q  \mathbb{M}'\}$. Now we are going to assume that $p_{\rm guess}^{\rm{DSD}}(\mathcal{G}, \mathbb{M}) \geq p_{\rm guess}^{\rm{DSD}}(\mathcal{G}, \mathbb{M}^*)$ holds for all games $\mathcal{G} = \{p(x, y), \sigma_{xy}\}$ and show show that there always exist a subroutine $\mathscr{S}$ which allows to simulate $\mathbb{M}^*$ using $\mathbb{M}$. We thus have:
\begin{align}
    \label{eq:1}
    \forall \, \mathcal{G} \qquad \max_{\mathbb{M}' \prec_q \mathbb{M}} \sum_{a,b} p(a, b) \, \tr\left[ M_{ab}' \, \sigma_{ab} \right] - \max_{\mathbb{M}'' \prec_q \mathbb{M}^*} \sum_{a,b} p(a, b) \, \tr\left[ M_{ab}'' \, \sigma_{ab} \right] \geq 0.
\end{align}
Let us now choose a particular subroutine in the second maximization, i.e.: $\mathscr{S}^* = \{p(\lambda) = \delta_{\lambda0},$ $p(a|i, \lambda) = \delta_{ai},$  $p(b|j, \lambda) = \delta_{bj}, \,U_{\lambda} = V_{\lambda} = \mathbb{1}\}$. In this way (\ref{eq:1}) implies:
\begin{align}
    \label{eq:2}
    \forall \, \mathcal{G} \qquad \max_{\mathbb{M}' \prec_q \mathbb{M}} \sum_{a,b} p(a, b) \, \tr\left[ (M_{ab}' - M_{ab}^*)\, \sigma_{ab} \right] \geq 0.
\end{align}
Let us denote $\Delta_{ab} := M_{ab}' - M_{ab}^*$. Since both $M_{ab}'$ and $M_{ab}^*$ are measurements we have that $\sum_{a,b} \Delta_{ab} = 0$. This also means that only one of the two situations can hold: either $(i)$ $\Delta_{ab} = 0$ for all $a, b$ or $(ii)$ there exists at least one $\Delta_{ab}$ with at least one negative eigenvalue. 

We will now show by contradiction that $(ii)$ cannot be true. Let us assume that $(ii)$ holds and label the negative eigenvalue with $\lambda_{a^*b^*}$ and the associated eigenvector with $\ket{\lambda_{a^*b^*}}$. Then, since (\ref{eq:2}) holds for all games $\mathcal{G}$, it also holds for a particular game $\mathcal{G}^* = \{p(a, b) = \delta_{aa^*}\delta_{bb^*}, \sigma_{ab} = \dyad{\lambda_{a^*b^*}}\}$. Hence (\ref{eq:2}) implies:
\begin{align}
    \bra{\lambda_{a^*b^*}} \Delta_{a^*b^*} \ket{\lambda_{a^*b^*}} = \lambda_{a^*b^*} < 0,
\end{align}
which is a contradiction. Hence we infer that $(ii)$ cannot be true and the only possibility is that each operator $\Delta_{ab}$ is identically zero. This means that:
\begin{align}
    M_{ab}^* = M_{ab}' := \sum_{i,j,\lambda} p(\lambda)p(a|i, \lambda) p(b|j, \lambda) (U_{\lambda}^{\dagger} \ot V_{\lambda}^{\dagger}) M_{ij} (U_{\lambda}^{} \ot V_{\lambda}^{}),
\end{align}
i.e. $\mathbb{M}^*$ can be simulated using $\mathbb{M}$.

\subsection{Proof of Result 7} 
\noindent The accessible min-information $I^{\rm acc}_{\rm min}(\mathcal{N})$ of a channel $\mathcal{N}$ is defined as \cite{Wilde2013}:
\begin{align}
    \label{eq:imax}
    I^{\rm {acc}}_{\rm min}(\mathcal{N}) = \max_{\mathscr{E}, \mathscr{D}}\,\, \big[ H_{\rm{min}}(X) - H_{\rm{min}}(X|G)\big],
\end{align}
where the optimization is over all encodings $\mathscr{E} = \{p(x), \sigma_x\}$ and decodings $\mathscr{D} = \{D_g\}$ and the min-entropies are defined as:
\begin{align}
    H_{\rm{min}}(X) &= -\log \max_x p(x),\\
    H_{\rm{min}}(X|G) &= - \log \left[\sum_{g}\max_{x}p(x,g)\right],
\end{align}
and $p(x, g)$ is the probability distribution induced by channel $\mathcal{N}$, i.e.:
\begin{align}
p(x, g) = p(x) p(g|x) = p(x) \tr\left[\mathcal{N}[\sigma_x] D_g \right].
\end{align}
Consider now encoding a bipartite random variable $X \times Y$ in an ensemble of bipartite quantum states, i.e.: $\mathscr{E} = \{p(x,y), \sigma_{xy}^{\rm AB}\}$ and $\mathscr{D} = \{D_g^{\rm A'B'}\}$ for $g = 1, \ldots, o_{\rm A}\cdot o_{\rm B}$. Moreover, consider the channel $\mathcal{N} = \mathcal{N}^{\rm AB \rightarrow A'B'}$ to be a quantum-to-classical measurement channel, which can be written as:
\begin{align}
    \mathcal{N}^{\rm AB \rightarrow A'B'}(\rho^{\rm AB}) = \sum_{a,b}\tr[M_{ab}^{\rm AB} \rho^{\rm AB}] \dyad{a}_{\rm A'} \ot \dyad{b}_{\rm B'},
\end{align}
where $\mathbb{M} = \{M_{ab}^{\rm AB}\}$ is a distributed measurement. We have:
\begin{align}
    I^{\rm {acc}}_{\rm min}(\mathcal{N}^{\rm AB \rightarrow A'B'}) &= \max_{\mathscr{E},\mathscr{D}} \log\left[\sum_{g} \max_{x,y} p(x,y) \tr[\mathcal{N}^{\rm AB \rightarrow A'B'}[\sigma_{xy}^{\rm AB}]D_g^{\rm A'B'}]  \right] - \log \max_{a,b} p(a,b) \\
    &= \max_{\mathscr{E},\mathscr{D}} \log\left[\sum_{g}\sum_{a,b} \max_{x,y} p(x,y) \tr[M_{ab}^{\rm AB}\sigma_{xy}^{\rm AB}] \tr[D_g^{\rm A'B'} \dyad{a}_{\rm A'}\ot\dyad{b}_{\rm B'}]  \right] - \log \max_{a,b} p(a,b) \\
    &= \log\left[\sum_{a,b} \max_{\mathscr{E}}  \max_{x,y} p(x,y) \tr[M_{ab}^{\rm AB}\sigma_{xy}^{\rm AB}]  \right] - \log \max_{a,b} p(a,b). 
    \label{eq:7}
\end{align}
Notice now that we can always express the optimization over $(x,y)$ as:
\begin{align}
    \max_{x,y} p(x,y) \tr[M_{ab}^{\rm AB}\sigma_{xy}^{\rm AB}] &= \max_{p(x|a)} \max_{p(y|b)} \sum_{x,y} p(x|a) p(y|b) p(x,y)  \tr[M_{ab}^{\rm AB}\sigma_{xy}^{\rm AB}] \\
    &= \max_{p(\lambda)}\max_{p(x|a, \lambda)} \max_{p(y|b, \lambda)} \sum_{x,y,\lambda} p(x|a,\lambda) p(y|b,\lambda) p(x,y)  \tr[M_{ab}^{\rm AB}\sigma_{xy}^{\rm AB}]
\end{align}
Notice further that if we carry out the optimisation of the above expression over $\mathscr{E}$ we can additionally write:
\begin{align}
   \max_{\mathcal{E}} \max_{x,y} p(x,y) \tr[M_{ab}^{\rm AB}\sigma_{xy}^{\rm AB}]  &= \max_{\mathscr{E}} \max_{p(\lambda)}\max_{p(x|a, \lambda)} \max_{p(y|b, \lambda)} \sum_{x,y,\lambda} p(x|a,\lambda) p(y|b,\lambda) p(x,y)  \tr[M_{ab}^{\rm AB}\sigma_{xy}^{\rm AB}] \\
   &= \max_{\mathscr{E}} \max_{\{\mathcal{E}_{\lambda}\},\{\mathcal{F}_{\lambda}\}} \max_{p(\lambda)}\max_{p(x|a, \lambda)} \max_{p(y|b, \lambda)} \sum_{x,y,\lambda} p(x|a,\lambda) p(y|b,\lambda) p(x,y) \times \\ 
   & \hspace{150pt} \tr[M_{ab}^{\rm AB} (\mathcal{E}_{\lambda}^{\rm A} \ot \mathcal{F}_{\lambda}^{\rm B})(\sigma_{xy}^{\rm AB})]
   \\
   &= \max_{\mathscr{E}} \max_{\mathbb{N} \prec \mathbb{M}} \sum_{x} p(x,y) \tr \left[N_{ab}^{\rm AB}\sigma_{xy}^{\rm AB}\right].
\end{align}
Hence we can further continue from (\ref{eq:7}) and write:
\begin{align}
    I^{\rm {acc}}_{\rm min}(\mathcal{N}^{\rm AB \rightarrow A'B'}) &= \log\left[\sum_{a,b} \max_{\mathscr{E}} \max_{\mathbb{N} \prec \mathbb{M}} p(a,b) \tr\left[N_{ab}^{\rm AB} \sigma_{ab}^{\rm AB}\right] \right] - \log \max_{a,b} p(a,b) \\
    &= \max_{\mathscr{E}} \ \log\left[ \max_{\mathbb{N} \prec \mathbb{M}} \sum_{a,b} p(a,b) \tr\left[M_{ab}^{\rm AB} \sigma_{ab}^{\rm AB}\right] \right] - \max_{a,b} p(a,b) \\
    &= \max_{\mathscr{E}} \log \left[ p_{\rm{guess}}^{\rm{DSD}}(\mathcal{G}, \mathbb{M}^{\rm AB})\right] - \log \left[p_{\rm{guess}}^{\rm DSD}(\mathcal{G})\right] \\
    &=  \log\left[\max_{\mathscr{E}} \frac{ p_{\rm{guess}}^{\rm{DSD}}(\mathcal{G}, \mathbb{M}^{\rm AB})}{p_{\rm{guess}}(\mathcal{G})} \right] \\
    &= \log \left[1 + \robn(\mathbb{M}^{\rm AB})\right].
\end{align}

\end{document}